\documentclass[12pt,preprint]{aastex}

\usepackage{xcolor}

\newcommand{\der}[2][\;\;]{\ensuremath{ \frac{d{#1}}{d{#2}} }}
\newcommand{\dern}[3][\;\;]{\ensuremath{ \frac{d^{#3}{#1}}{d{#2}^{#3}} }}
\newcommand{\dpar}[2][\;\;]{\ensuremath{ \frac{\partial{#1}}{\partial{#2}} }}
\newcommand{\dparn}[3][\;\;]{\ensuremath{ \frac{\partial^{#3}{#1}}{\partial{#2}^{#3}} }}

\newcommand{\bvec}[1]{{\mbox{{\boldmath$#1$}}}}		
\newcommand{\unitv}[1]{\bvec{\hat{#1}}}			
\newcommand{\grad}{\bvec{\nabla}}			
\newcommand{\va}{V_{\rm A}}				
\newcommand{\oma}{\omega_{\rm A}}			
\newcommand{\ra}{r_{\rm A}}				
\newcommand{\vsep}{\, | \, }

\newcommand{\phfl}{Phys. Fluids}

\newcommand{\cpc}{Comput. Phys. Commun.}

\newcommand{\eqnref}[1]{(\ref{#1})}

\begin{document}

\title{A Novel Approach to Resonant Absorption of the Fast MHD Eigenmodes of a Coronal Arcade}
\shorttitle{Resonant Absorption of the Eigenmodes of an Arcade}

\author{Bradley W. Hindman}
\affil{JILA, NIST and University of Colorado, Boulder, CO~80309-0440, USA}

\author{Rekha Jain}
\affil{School of Mathematics \& Statistics, University of Sheffield, Sheffield S3 7RH, UK}

\email{hindman@solarz.colorado.edu}
\today


\begin{abstract}

\end{abstract}

The arched field lines forming coronal arcades are often observed to undulate as
magnetohydrodynamic (MHD) waves propagate both across and along the magnetic field.
These waves are most likely a combination of resonantly coupled fast magnetoacoustic
waves and Alfv\'en waves. The coupling results in resonant absorption of the fast 
waves, converting fast wave energy into Alfv\'en waves. The fast eigenmodes of the
arcade have proven difficult to compute or derive analytically, largely because of
the mathematical complexity that the coupling introduces. When a traditional spectral
decomposition is employed, the discrete spectrum associated with the fast eigenmodes
is often subsumed into the continuous Alfv\'en spectrum. Thus fast eigenmodes,
become collective modes or quasi-modes. Here we present a spectral decomposition
that treats the eigenmodes as having real frequencies but complex wavenumbers.
Using this procedure we derive dispersion relations, spatial damping rates, and
eigenfunctions for the resonant, fast eigenmodes of the arcade. We demonstrate
that resonant absorption introduces a fast mode that would not exist otherwise.
This new mode is heavily damped by resonant absorption, only travelling a few
wavelengths before losing most of its energy.



\section{Introduction}
\label{sec:Introduction}

One of the most prominent features seen in images of the solar corona by EUV
telescopes are the elegant arches of glowing plasma that trace magnetic field
lines through the corona. Typically, these loops are preferentially illuminated
segments of a larger magnetic structure comprised of an arcade of arched field
lines. Coronal loops and arcades are full of waves. They are often observed to
shimmer and undulate, sometimes in clear response to nearby solar flares
\citep[e.g.,][]{Aschwanden:1999, Nakariakov:1999, Wills-Davey:1999} but also
often without an obvious, visible excitation event \citep[e.g.,][]{Nistico:2013,
Anfinogentov:2013, Duckenfield:2018}. Typically, oscillations generated by
flares and other impulsive events have an amplitude that suddenly rises and
then decays rapidly over a handful of wave periods \citep[e.g.,][]{White:2012,
Goddard:2016}. Conversely, ambient oscillations that lack an obvious source
event (often called ``decayless" oscillations) are usually of low amplitude
and oscillate for long durations without significant attenuation of the signal.

The commonly accepted view is that coronal loops are long tube-like magnetic
structures and their oscillations are caused by MHD kink waves that are confined
to the loop and trapped between the two footpoints where the loop intersects
the photosphere \citep[see the review by][]{Andries:2009}. This model has
the useful feature that the wave propagation and trapping can be reduced to
a 1-D wave problem. The rapid attenuation of the wave signal that is observed
for many loop oscillations has been explained using a variety of mechanisms,
with resonant absorption being the most prominent \citep[e.g.,][]{Ruderman:2002,
Goossens:2002, Goossens:2011}. Resonant absorption is actually a mode conversion
mechanism instead of true dissipation. The fast kink wave resonantly couples
to a local Alfv\'en wave at locations where the two wave modes share a common
frequency \citep[e.g.,][]{Chen:1974a, Chen:1974b, Goedbloed:1971}. In ideal
MHD this energy transformation occurs on infinitely thin critical surfaces
where the MHD equations formally become singular. The Alfv\'en waves are thus
highly localized and rapidly dissipate when nonideal effects are included.

\cite{Hindman:2014} recently suggested that the rapid diminuation of the
loop-oscillation amplitude may naturally result from the transit of a fast
MHD wave packet propagating down the arcade in which the loop is embedded. In
this model, the entire arcade participates in the oscillation but since the
coronal loop is preferentially bright, the motion of those special field lines
are particularly obvious. If this suggestion is correct, the rapid reduction
in wave amplitude that is observed is not due to dissipation or loss of fast
wave energy. Instead, the profile of the time series depends on the evolving
shape of the wave packet as it passes by the visible loop when propagating
down the axis of the arcade.

Magnetic arcades have long been known to form waveguides. Explicitly, for a
coronal arcade, fast waves are trapped from below by reflection from the photosphere's
large mass density and from above by refraction by the ever increasing Alfv\'en
speed with height. In the horizontal direction aligned with the axis of the
arcade, trapping is likely to be only partial and for much of the arcade, the
waves can propagate freely in the axial direction. Similar magnetic structures
have been explored in the context of the Earth's magnetosphere
\citep[e.g.,][]{Southwood:1974, Kivelson:1985, Kivelson:1986} and in fusion
devices in the form of Z-pinches. From such considerations, one would expect
a discrete spectrum of fast wave eigenmodes with eigenfrequencies that depend
on one continuous wavenumber and two quantized wavenumbers. The continuous
wavenumber corresponds to variation in the direction binormal to the field lines,
i.e., parallel to the arcade's axis. The two quantized wavenumbers are in the
directions parallel to the field and in the principal normal to the field lines.

Numerical models of MHD wave propagation have identified such fast wave resonances
\citep[e.g.,][]{Oliver:1998, Arregui:2004, Rial:2010, Rial:2013} through the
enhanced response to a driver. Insufficient attention has previously been paid
to the analytic calculation of the eigenspectrum of arcades for the simple
reason that the mathematics is suprisingly challenging. In such a magnetic geometry
the local Alfv\'en frequency is in general a function of position. Thus, there
is a continuous spectrum of allowed Alfv\'en waves \citep[e.g.,][]{Goossens:1985,
Poedts:1985, Poedts:1988}. The fast waves can resonantly couple to the Alfv\'en
waves leading to resonant absorption. In this case, for a given frequency the
singularity corresponds to one or more curved flux surfaces. Detailed ideal MHD
calculations of the {\bf eigenfrequencies} demonstrate that the expected discrete
spectrum of fast modes is subsumed into the continuous spectrum associated with
the Alfv\'en continuum \citep[e.g.,][]{Goedbloed:1971, Lee:1986, Goedbloed:2004}.
The fast wave eigenmodes are therefore quasi-modes or collective modes. Calculations
that include physical dissipation (resistivity, viscosity, etc.) reveal that
that these quasi-modes correspond to true eigenmodes of the dissipative
spectrum \citep[e.g.,][]{Kerner:1985, Kerner:1986, Pao:1985, Poedts:1991}, albeit with
eigenfunctions that do not converge to the ideal eigenfunctions when the limit
of zero dissipation is considered.

Much of the theoretical work on waves within 3D models of coronal arcades
have avoided these critical layers by inserting artificial boundaries that
exclude the resonant field lines. Further, the geometry has often been simplified
by focusing on Cartesian slabs of magnetized plasma and building on the initial
model of \cite{Edwin:1982}. A few studies have explored the effects of field-line
curvature by examining warped slabs with either cylindrical \citep{Smith:1997,
Brady:2005, Selwa:2005, Verwichte:2006a, Verwichte:2006b} or elliptical geometry
\citep{Diaz:2006}. All of these studies have assumed that the wave cavity is radially
confined to the shell formed by the visible bright loops and in doing so, artificial
radial boundaries are imposed. Despite these boundaries, all find that radial
cross-field wave leakage couples the shell or slab to the rest of the corona.
Alternatively, \cite{Hindman:2015} and \cite{Thackray:2017} explored analytic
solutions to waves propagating within semi-infinite atmospheres. These two studies
considered radial Alfv\'en speed profiles that naturally trap waves without the
need for artificial boundaries or discontinuities. However, they\ avoided the
critical layers by intentionally choosing piece-wise constant profiles for the
local Alfv\'en frequency.

Our goal here is to deal with the critical layers directly. We will compute the
eigenspectrum for a smooth but spatially varying Alfv\'en speed profile, examining
both the dispersion relation for the fast wave eigenmodes and the damping efficiency
of the resonant absorption. The merger of the discrete fast wave spectrum with the
continuous Alfv\'en spectrum will be avoided by assuming translational invariance
of the background magnetic field in the axial direction and employing a {\bf nonstandard}
spectral decomposition. Traditionally, the eigenvalues are considered to be
eigenfrequencies. But, when the background magnetic field is axially invariant, we
are allowed to treat the axial wavenumber as the eigenvalue and the frequency as
a continuous parameter. This switch in what is considered the eigenvalue results
in a separation of the discrete fast wave spectrum and the Alfv\'en continuum
even though the two wave modes remain resonantly coupled. Thus, the fast
wave eigenfunctions are well-defined and can be computed though straightforward
semi-analytic techniques. The calculation of these eigenfunctions is a necessary
first step toward constructing wave packets of fast waves that can propagate down
the arcade and appear as loop oscillations as they pass by bright bundles of field
lines.

This paper has the following layout.  In section~\ref{sec:ArcadeModel} we present
a cylindrical model of a coronal arcade. In section~\ref{sec:GoverningEqns} we
derive an ODE that describes the radial behavior of the fast waves for our cylindrical
arcade. We solve for the discrete spectrum of fast wave eigenmodes in section~\ref{sec:Eigenspectrum},
where we present dispersion relations and eigenfunctions. Finally, in section~\ref{sec:Discussion}
we discuss the nature of the modes that we have obtained and explore the implications
for observations of coronal loop oscillations.


\section{A cylindrical model of a coronal arcade}
\label{sec:ArcadeModel}

We adopt a simple model of a coronal arcade that remains tractable while still
including the important effects of inhomogeneity and field-line curvature. We
assume that the corona is magnetically dominated (i.e., the ratio of gas pressure
to magnetic pressure is small, $\beta <<1$) and consider a potential magnetic field
generated by a line current of strength $I$ embedded in the solar photosphere. We
employ a cylindrical coordinate system ($r$, $\theta$, $y$), where the coordinate
axis, $\unitv{y}$, is collinear with the line current. The photosphere is a flat
plane and the line current is a straight line that lies within that plane. Each
field line is a semi-circle with the magnetic field pointing purely in the azimuthal
direction, $\unitv{\theta}$. The two ends of each field line (located at $\theta = 0$
and $\theta = \pi$) are anchored in the photosphere. The field strength, $B$, is
a function of only the cylindrical radius, $r$,

\begin{equation} \label{eqn:potential_field}
     \bvec{B} = B(r) \, \unitv{\theta} = \frac{2I}{r} \, \unitv{\theta}\; .	
\end{equation}

\noindent The field is therefore axisymmetric, with spatial dependence in radius,
$r$, and invariant along its axis, $y$. We will generate MHD wave solutions in
the half-space lying above the photosphere,  i.e., $\theta \in [0,\pi]$,
$r \in[0,\infty)$, and $y \in(-\infty,\infty)$.

In order to ensure that separable solutions exist, we assume that the mass density
is a function of cylindrical radius alone, $\rho=\rho(r)$. Given the geometry of
the magnetic field, this assumption dictates that the Alfv\'en speed, $\va(r) = B/\sqrt{4 \pi \rho}$,
is also solely a function of radius. This density  variation is consistent with
hydrostatic balance along field lines as long as the corona is exceedingly hot
such that the density scale height due to gravitation is  much larger than the
height of the oscillating loops in the arcade.

When choosing an Alfv\'en speed profile, we must place two restrictions: (1)
The Alfv\'en speed approaches a nonzero value, $V_0$, near the axis ($r=0$)
and (2) the Alfv\'en speed increases monotonically with radius beyond a fiducial
distance from the coordinate axis. The first of these conditions ensures that
the solution is {\bf not} recessive at the axis---i.e., the solutions have a
nonzero radial wavelength \citep[see][]{Hindman:2015}. The second condition
guarantees that a cylindrical waveguide exists \citep[e.g.,][]{Terradas:1999}
by arranging that outward propagating waves are refracted back inward at an
outer turning point. We adopt the following Alfv\'en speed profile which
satisfies both of these conditions in a simple manner,

\begin{equation} \label{eqn:VaProfile}
	\va^2(r) = V_0^2 \left(e^{-r^2/r_0^2} + \frac{r^2}{r_0^2}\right) \; .
\end{equation}

\noindent In this profile, $r_0$ and $V_0$, are arbitrary constants that represent
a characteristic scale length and the minimum speed achieved at the coordinate
axis. When needed these two constants will be used to nondimensionalize all variables.
Figure~\ref{fig:AlfvenProfile}a shows this Alfv\'en speed profile as a function
of cylindrical radius. The functional form of this profile is monotonic with radius
and as the radius becomes large, $r>>r_0$, it {\bf rapidly} approaches a linear
function of radius. These properties allow the analytic solutions of \cite{Hindman:2015}
to be used as numerical boundary conditions for large radius.


\section{Governing Wave Equation}
\label{sec:GoverningEqns}

Since the plasma is magnetically dominated, we can safely ignore gas
pressure and buoyancy in the equation of motion.  Under this ``cold-plasma"
approximation, the wave motions become purely transverse to the magnetic field
because the only remaining force, the Lorentz force, is itself transverse. Therefore,
the azimuthal component of the fluid's velocity vector, $u_\theta$, is identically
zero, and only the radial and axial components need be considered,
$\bvec{u} = u_r \, \unitv{r} + u_y \, \unitv{y}$. For such transverse motions,
the linearized MHD induction equation dictates that the fluctuating magnetic
field $\bvec{b}$ is as follows:

\begin{eqnarray} \label{eqn:induction}
	\dpar[\bvec{b}]{t} &=& \frac{B}{r} \dpar[u_r]{\theta} \, \unitv{r}
		+ B \Phi \, \unitv{\theta} + \frac{B}{r} \dpar[u_y]{\theta} \, \unitv{y} \; ,
\\
	\label{eqn:Def_Phi}
	\Phi &\equiv& -\left(\frac{r}{B}\right) \, \grad_\perp \cdot \left(\frac{B\bvec{u}}{r}\right)
		= -r^2 \, \grad_\perp \cdot \left(\frac{\bvec{u}}{r^2}\right) \; ,
\end{eqnarray}

\noindent where $\grad_\perp$ is the component of the gradient operator that is transverse
to the background magnetic field,

\begin{equation}
	\grad_\perp \equiv \unitv{r} \dpar[]{r} + \unitv{y} \dpar[]{y} \; .
\end{equation}

\noindent The variable $\Phi$ is proportional to the temporal derivative of
the fractional magnetic-pressure fluctuation,

\begin{equation} \label{eqn:pressure_fluctuation}
	\dpar[]{t}\left(\frac{\bvec{B} \cdot \bvec{b}}{4 \pi}\right)
		= \frac{B^2}{4 \pi} \Phi \; .
\end{equation}

For the magnetically dominated plasma discussed previously, the linearized MHD
momentum equation takes on a relatively simple form,

\begin{equation} \label{eqn:vector_MHD}
	\dparn[\bvec{u}]{t}{2} = \frac{\va^2}{r^2}
		\left(\dparn[u_r]{\theta}{2} \, \unitv{r}
			+ \dparn[u_y]{\theta}{2} \, \unitv{y} \right)
		- \va^2 \grad_\perp \Phi \; ,
\end{equation}

\noindent In equation~\eqnref{eqn:vector_MHD} the term involving $\grad_\perp\Phi$
represents the transverse components of the magnetic-pressure force and the two
terms in parentheses comprise the transverse components of the force generated
by the magnetic tension. Equation~\eqnref{eqn:vector_MHD} is a coupled set of
PDEs that describes both Alfv\'en waves and fast magnetoacoustic waves. The slow
waves have been removed by our low-$\beta$ approximation. Since the fast waves
are magnetic-pressure waves, we can derive an equation for the fast waves by seeking
an equation with $\Phi$ as the sole independent variable.

We reduce the PDEs to ODEs by exploiting the symmetries of the arcade. We
remind the reader that the arcade is invariant in the axial and azimuthal directions.
We assume that the arcade is sufficiently long in the axial $y$-direction that
we can ignore boundary effects. Further, we enforce a line-tying boundary condition
(i.e., stationary field lines) at the photosphere. With these assumptions, it
is convenient to perform Fourier transforms in time $t$ and in the axial spatial
coordinate $y$ and to perform a sine series expansion in azimuth $\theta$. We
choose our Fourier conventions such that each wave component with a unique
combination of temporal frequency $\omega$, axial wavenumber $k$, and azimuthal
order $m$ has the following functional form:

\begin{equation} \label{eqn:mode_form}
	\Phi(r,\theta,y,t) \to \Phi_m(r \vsep k,\omega) \; \sin\left(m \theta\right) \; e^{i\left(ky-\omega t\right)} \; .
\end{equation}

After transformation, equation~\eqnref{eqn:vector_MHD} becomes,

\begin{equation} \label{eqn:u_ODE}
	\left(\frac{\omega^2 - \oma^2}{\va^2}\right) \bvec{u}_m = \grad_\perp \Phi_m \; ,
\end{equation}

\noindent where we define the local Alfv\'en frequency $\oma$,

\begin{equation}
	\oma(r) \equiv \frac{m \, \va(r)}{r} \; .
\end{equation} 

\noindent Figure~\ref{fig:AlfvenProfile}b presents the Alfv\'en frequency
as a function of radius for the Alfv\'en speed profile~\eqnref{eqn:VaProfile}
used in our numerical calculations. For display purposes, the frequency is
shown for an azimuthal order of unity, $m=1$.

For compactness of notation we henceforth drop the subscript $m$ whenever doing
so would not result in confusion. Further, we use the same symbol for a variable
in physical space and in spectral space; only the arguments (and context) distinguish
one from the other. A single second-order ODE can be obtained in the variable
$\Phi$ by taking the transverse divergence of equation~\eqnref{eqn:u_ODE} and
using the definition of the fractional magnetic-pressure fluctuation,
equation~\eqnref{eqn:Def_Phi}, to eliminate the velocity,

\begin{equation} \label{eqn:PhiODE}
	r K^2 \der[]{r}\left(\frac{1}{r K^2} \der[\Phi]{r}\right) +
		\left(\frac{\omega^2}{\va^2}-\frac{m^2}{r^2} - k^2\right) \Phi = 0 \; .
\end{equation}

\noindent In this equation, $K$ is a local wavenumber,

\begin{equation}
	K^2(r) \equiv \frac{\omega^2-\oma^2(r)}{\va^2(r)} \, ,
\end{equation}

\noindent that vanishes at the Alfv\'en resonances. Equation~\eqnref{eqn:PhiODE}
has the same essential singularities as the Generalized Hain-L\"ust equation
\citep{Hain:1958, Goedbloed:1971} when that equation is considered in the limits
of low-$\beta$ and vanishing axial field ($\bvec{B} \, \cdot \, \unitv{y} = 0$).
Our equation is, however, rather simpler in form since it lacks the ``apparent"
singularities that arise from fast-mode turning points \citep[see][]{Appert:1974,
Goedbloed:2004}. The Hain-L\"ust equation is inherently more complicated since
it describes the radial variation of the radial velocity component $u_r$, which
contains direct contributions from all three MHD wave modes (fast, slow, and
Alfv\'en). Our equation for the fractional magnetic-pressure fluctuation has
a direct contribution from only the fast mode.

For a general Alfv\'en speed profile, $\va(r)$, equation~\eqnref{eqn:PhiODE}
has internal critical surfaces located at the Alfv\'en resonances, i.e., at the
radii $r=\ra$ such that $\oma^2(\ra) = \omega^2$ or equivalently $K^2(\ra) = 0$.
For the monotonic Alfv\'en speed profile given by equation~\eqnref{eqn:VaProfile},
there is only one resonant radius for any given frequency and that resonant radius
can be expressed in terms of Lambert $W$-functions, 

\begin{equation}
	\ra^2 = r_0^2 \, W\left( -\frac{m^2}{\nu^2} \right) \; ,
\end{equation}

\noindent with $W(x)$ being the solution to Lambert's transcendental equation
$W e^W = x$ \citep{Corless:1996} and $\nu$ being a frequency-dependent parameter
defined by $\nu^2 \equiv m^2 - \omega^2 r_0^2 V_0^{-2}$.

Mathematically, these critical surfaces correspond to logarithmic regular singular
points of the ODE~\eqnref{eqn:PhiODE} \citep[e.g.,][]{Uberoi:1972, Appert:1974,
Goedbloed:1971, Goedbloed:1975}. Such singularities result in the resonant absorption
of fast waves through coupling to the continuum of possible Alfv\'en waves
\citep[e.g.,][]{Hollweg:1990, Wright:1992, Goedbloed:2004}. In \cite{Hindman:2015}
these critical cylindrical surfaces were circumvented by choosing an Alfv\'en frequency
profile that was piece-wise constant with radius. Here we will examine a more general
profile and therefore we must deal with the singularity explicitly.


\subsection{Structure of the Wave Cavity}

A WKB analysis of equation~\eqnref{eqn:PhiODE} reveals that waves of all frequencies
are trapped as long as the wave is obliquely propagating, i.e., $k\neq 0$. By making
the transformation $\Phi = r^{1/2} K \Psi$, our ODE~\eqnref{eqn:PhiODE} is converted
into a standard Helmholtz equation,

\begin{equation} \label{eqn:WKB_form}
	\dern[\Psi]{r}{2} + \left[\frac{\omega^2 - \omega_s^2}{\va^2} - \frac{m^2}{r^2} -k^2\right] \; \Psi = 0 \,
\end{equation}

\noindent with a ``cut-off" frequency,

\begin{equation} \label{eqn:Cutoff_Frequency}
	\omega_s^2 \equiv r^{1/2}K\va^2  \dern[ ]{r}{2} \left(\frac{1}{r^{1/2}K}\right) \; ,
\end{equation}

\noindent that arises from the Alfv\'en singularity and is itself a function of
frequency (through $K$). Assuming real frequencies and axial wavenumbers, the
wave cavity spans those radii where $\omega^2 > \omega_s^2 +\left(m^2 r^{-2} + k^2\right) \va^2$.

\begin{figure*}
	\epsscale{1.0}
	\plotone{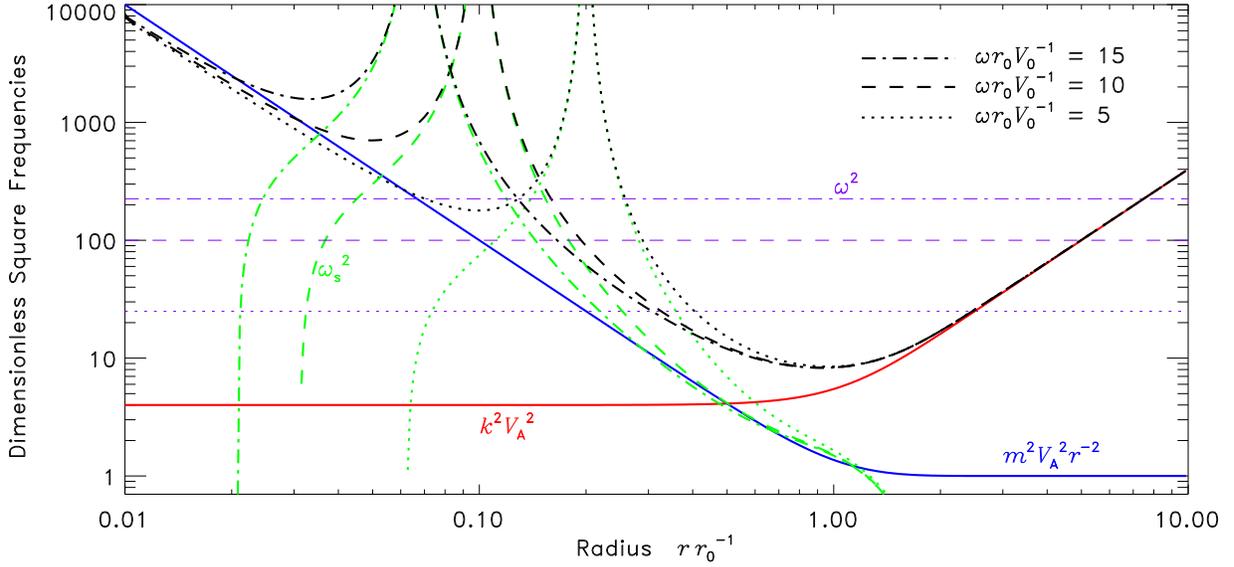}
	\caption{\small A propagation diagram that illustrates the radial cavity
for the fast modes. The wave cavity exists wherever the waves are radial propagating,
which corresponds to the range of radii where the wave frequency exceeds a critical
frequency constructed by adding three special frequencies in quadrature
$\omega^2 > k^2 \va^2 + m^2 \va^2 \, r^{-2} + \omega_s^2$. The red curve corresponds
to $k^2 \va^2$ and the blue curve indicates $m^2 \va^2 \, r^{-2}$. The green curves
show the square of the singular cut-off frequency, $\omega_s^2$, for the three
dimensionless frequencies, $\omega r_0 V_0^{-1} = 5$, 10, and 15. See
equation~\eqnref{eqn:Cutoff_Frequency} for a definition of the cut-off frequency.
The line style used for each frequency is indicated in the upper-right of the figure.
The black curves indicate the critical frequencies obtained by adding the red, blue
and green curves together. The squares of the three frequencies used to generate the
cut-offs are indicated with the horizontal violet lines.  All frequencies are generated
for unit azimuthal order, $m=1$.
	\label{fig:Propagation_Diagram}}
\end{figure*}

Figure~\ref{fig:Propagation_Diagram} illustrates the special frequencies that
define the wave cavity. All frequencies are nondimensionalized by $V_0 r_0^{-1}$
and plotted in the form of their squares. The red curve corresponds to the quantity
$k^2 \va^2$ for an axial wavenumber of $k r_0 = 2$. The blue curve shows the square
of the local Alfv\'en frequency $\omega_A^2 = m^2 \va^2 /r^2$, which in these equations
arises from the azimuthal wavenumber $m/r$ (and {\bf not} from the Alfv\'en singularity).
The three green curves represent the singular cut-off frequency for three different
dimensionless frequencies, $\omega r_0 V_0^{-1} = [5, 10, 15]$. The lowest frequency
is shown by the dotted curve, the intermediate frequency by the dashed curve,
and the highest frequency by the dot-dashed curve. The squares of the three wave
frequencies are indicated with the horizontal violet lines using the same line
styles. Finally, the critical frequency given by the sum of all three special
frequencies, $\omega_s^2 + \omega_A^2 + k^2 \va^2$, is shown using black curves.
The cavity exists wherever the wave frequency (violet) exceeds the critical
frequency (black).

From Figure~\ref{fig:Propagation_Diagram} one can easily deduce that there is a
single fast-wave cavity, bracketed by an inner and outer turning point. The outer
turning point, $r_2$, is refractive and occurs where $\omega = k \va$. For the
relatively high frequencies shown, the Alfv\'en speed is nearly a linear function
of radius at the outer point and the outer turning point is therefore proportional
to the axial phase speed of the wave,

\begin{equation}
	r_2 \to \frac{r_0}{V_0} \, \frac{\omega}{k}  \; .
\end{equation}

From this expression, we obtain the expected result that only waves that
propagate at least partially in the axial direction are trapped. Purely radially
propagating waves ($k=0$) are not refracted and therefore travel off to infinity.

The behavior of the inner turning point is more complicated. In the absence of the
singularity, the fast wave would be excluded from the cylindrical origin by the
geometrical effect of the azimuthal wavenumber, $m/r$ (the blue curve). However,
in the presence of the singularity, the singular term dominates $\omega_s^2 >> m^2 \va^2/r^2$,
and the inner turning point is moved outward and is largely dependent on the singularity.
This means that the singularity occurs outside the fast-wave cavity, within the
inner evanescence zone.  Therefore, we should expect that the fast modes are
inefficiently coupled to the Alfv\'en waves and only undergo weak damping.  We will
find that this expectation holds except for one noted exception.

An estimate for the inner turning point, $r_1$, can be obtained in the high-frequency
limit by examining the behavior of the singular cut-off near resonance. For high frequencies,
the Alfv\'en singularity approaches the origin. This can be verified by keeping only
the first term in the power series expansion of the Lambert $W$-function,
$W(x) = x - x^2 + (3/2) x^3 + \cdots$ to show that the Alfv\'en radius approaches

\begin{equation}
	r_A \to m \, \frac{V_0}{\omega} \; .
\end{equation}

\noindent In this limit, near the singularity, the square of the cut-off frequency
is divergent and positive,

\begin{equation}
	\omega_s^2 \to \frac{1}{4} \frac{V_0^2}{\left(r-\ra\right)^2} \; .
\end{equation}

\noindent By combining the previous two equations and setting $\omega = \omega_s$,
we find that the inner turning point is inversely proportional to frequency for
high frequencies,

\begin{equation}
	r_1 \to \left(m+2\right) \frac{V_0}{\omega} \; .
\end{equation}


\section{Eigenspectrum}
\label{sec:Eigenspectrum}

The eigenfunctions of equation~\eqnref{eqn:PhiODE} are those solutions that satisfy
boundary conditions of regularity at the coordinate axis ($r = 0$) and at infinity
($r\to\infty$). Appendix~\ref{sec:Singularities} provides a detailed discussion of
the nature of the solutions at both boundaries. In short, the inner solution $\Phi_0$,
i.e., the solution that is regular at the origin, behaves like a power law near the
axis, $\Phi_0 \sim r^m$. The outer solution, which is regular at infinity, behaves
asymptotically like a modified Bessel function of the second kind, $\Phi_\infty \sim K_\nu(kr)$,
with a frequency-dependent order, $\nu^2 \equiv m^2 - \omega^2 r_0^2 V_0^{-2}$. The
eigenfunctions are the solutions that satisfy both regularity conditions and correspond
to those frequencies and wavenumbers where the inner and outer solutions become linearly
dependent (i.e., their Wronskian vanishes). After satisfying both boundary conditions,
the eigenfunctions lack the  necessary free parameters to also satisfy a regularity
condition at the internal singularity associated with the Alfv\'en resonance $r=\ra$.
Hence, the solution is perforce irregular on the resonant field line.

In Appendix~\ref{subsec:interior}, we present detailed derivations of the two solutions
valid near the Alfv\'en resonances. Here we simply point out that one of the solutions
is irregular and possesses a logarithmic singularity. A careful analysis of the irregular
solution reveals that $\Phi$ itself is finite at the singularity, but the corresponding
velocity field---obtained through equation~\eqnref{eqn:u_ODE}---is discontinuous and
singular in both the radial and axial components. This poor behavior should be expected.
A fast wave with a purely monochromatic frequency can be treated as if the wave has
been resonantly transferring energy to the resonant field line for an infinite duration
of time.

The eigenmodes correspond to the subset of frequencies and wavenumbers for which the
Wronskian of the inner and outer solution vanishes,

\begin{equation}
	{\cal W}\left\{\Phi_0,\Phi_\infty\right\}\left(r\right) =
		\Phi_0 \der[\Phi_\infty]{r} - \Phi_\infty \der[\Phi_0]{r} = 0 \; .
\end{equation}	

\noindent Since our ODE has Sturm-Liouville form (albeit with a singular weight function),
a more convenient description of the modal condition involves the dispersion function $D$
which is proportional to the Wronskian,

\begin{equation}
	D\left(k,\omega\right) \equiv \frac{\cal{W}}{rK^2} = \frac{1}{r K^2} \left(\Phi_0 \, \der[\Phi_\infty]{r} - \der[\Phi_0]{r} \, \Phi_\infty \right) = 0\; .
\end{equation}

\noindent The dispersion function, $D(k,\omega)$, has the useful property that it
is independent of radius. This can be verified through direct differentiation,
$dD/dr=0$, and use of the ODE~\eqnref{eqn:PhiODE}.

In systems without damping, the modal condition $D(k,\omega) = 0$ is rather
straightforward. There are often a countable infinity of real zeros and one can
consider either the frequency $\omega$ or the axial wavenumber $k$ as a real free
parameter. Traditionally, one usually treats the wavenumber $k$ as the free parameter
and the eigenvalues are discrete eigenfrequencies $\omega_{mn}(k)$ where the radial
order $n$ is an integer that describes the number of nodes in the radial eigenfunction.
Note, however, that it is equally valid to consider the frequency as the free parameter
and the eigenvalues are eigenwavenumbers, $k_{mn}(\omega)$.

When the modes are damped, the eigenvalues becomes complex. Once again, it is
traditional to think in terms of a continuous real wavenumber with a complex
eigenfrequency $\omega_n(k) = \varpi_n(k) + i \gamma_n(k)$, where the dependence
on the azimuthal order $m$ has been dropped from the notation. The real part
of the eigenfrequency is inversely proportional to the wave period $P_n = 2\pi/\varpi_n$,
and the imaginary part defines the temporal damping rate $\gamma_n$. But as
before, it is equally valid to consider real frequencies and complex eigenwavenumbers,
$k_n(\omega) = \kappa_n(\omega) + i \eta_n(\omega)$. Here, the real part of
the complex wavenumber is related to the wavelength $\lambda_n = 2\pi/\kappa_n$
and the imaginary part is the {\bf spatial} decay rate, $\eta_n$. Usually it
does not matter which viewpoint one adopts; however, in this case the perverse
nature of the internal singularity actually makes the latter view more fruitful.
We discuss this issue in more detail in Appendix~\ref{sec:ContSpectra}.


\subsection{Numerical Solutions}
\label{subsec:numerical_solutions}

We numerically solve for the inner and outer solutions (and hence the dispersion
function) by using a somewhat complicated shooting technique. For the inner solution
$\Phi_0$ we start our numerical integrations at a small radius, $r_{\rm in} << \inf(r_0,\ra)$,
and obtain $\Phi_0(r_{\rm in})$ and its derivative by using a power-series expansion
that is valid near the origin (see Appendix~\ref{subsec:origin}). Then, using a
5th-order, adaptive-stepsize, Runge-Kutta algorithm, we numerically integrate
equation~\eqnref{eqn:PhiODE} from $r_{\rm in}$ to any chosen radius $r=R$. When
evaluating the dispersion function the actual value of $R$ that is chosen is
immaterial since the dispersion function is independent of radius. If $R < \ra$,
this desired radius lies inside the Alfv\'en resonance and we do not need to
explicitly worry about the interior singularity. However, if $R > \ra$ we need to
integrate through the singularity and this, of course, requires some care.

When the singularity lies between the origin and the desired radius $R$, we match
the inner solution to a linear combination of the solutions that are regular
$\Phi_{\rm reg}$ and irregular $\Phi_{\rm irr}$ at the Alfv\'en singularity,

\begin{equation} \label{eqn:Match_Phi_Inner}
	\Phi = A_{\rm reg} \Phi_{\rm reg} + A_{\rm irr} \Phi_{\rm irr} \; .
\end{equation}

\noindent This is accomplished by exploiting power-series expansions for $\Phi_{\rm reg}$
and $\Phi_{\rm irr}$ that are valid near the interior singular point (see
Appendix~\ref{subsec:interior}). Each expansion is evaluated just inside the singularity,
i.e., at $r = r_A - \delta$ where $\delta$ is positive and small. This provides
a starting point for the numerical integrations. Then using the same Runge-Kutta
integrator, we numerically integrate both solutions from $r = r_A - \delta$ to a
point half-way between the origin and the Alfv\'en singularity, $r = \ra/2$. The
inner solution $\Phi_0$ is then integrated from $r_{\rm in}$ near the origin to
the same point. The connection coefficients, $A_{\rm reg}$ and $A_{\rm irr}$, are
chosen to ensure that the inner solution $\Phi_0$ and the solution initially generated
near the singularity match cleanly (their functions and their derivatives match).
Once computed these connection coefficients can subsequently be used to generate
the inner solution on the far side of the singularity by using the power-series
expansions to evaluate $\Phi_{\rm reg}$ and $\Phi_{\rm irr}$ at $r = r_A + \delta$.
This solution (comprised of the linear combination of regular and irregular solutions)
is then numerically integrated to the desired radius $R$. We use a similar procedure
to numerically evaluate the outer solution. The only difference is that the starting
radius is large, $r_{\rm out} >> \sup(r_0,\ra)$, and an asymptotic solution that
is valid for large radius is used to initiate the numerical integrations (see
Appendix~\ref{subsec:infinity}).


\subsection{Dispersion Function}
\label{subsec:dispersion_function}
 
\begin{figure*}
	\epsscale{0.5}
	\plotone{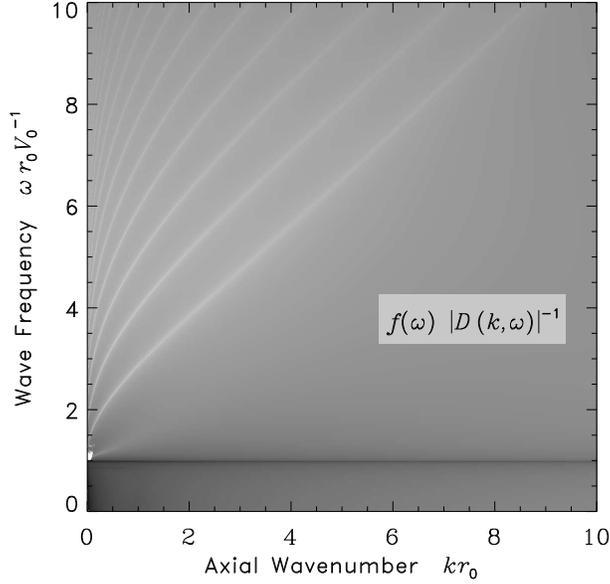}
	\caption{\small The reciprocal of the modulus of the dispersion function
multiplied by a function of frequency, $f(\omega) \, |D(k,\omega)|^{-1}$, is displayed
as a function of real dimensionless wavenumber $kr_0$ and frequency $\omega r_0 V_0^{-1}$.
The azimuthal order is unity, $m = 1$. The function of frequency was chosen solely
to remove a strong gradient in frequency from the image, $f(\omega) = \exp\left(-0.2 \; \omega r_0 V_0^{-1}\right)$.
The eigenmodes correspond to the bright ridges where the dispersion function is
nearly zero. The dark horizontal band lies at a frequency $\omega = m V_0/r_0$ which
corresponds to the frequency where the Alfv\'en resonance has moved to an infinite
radius. All eigenmodes exist above this frequency bound. The lowest frequency mode
(i.e., the ridge with the lowest radial order) is heavily damped and visible only
at low wavenumbers ($kr_0 < 1$). At higher wavenumbers, this mode's line profile
becomes sufficiently wide that the mode disappears into the background. The sharp-edged,
bright feature near $\omega r_0 V_0^{-1} = m$ at very low wavenumber ($kr_0 < 0.1$)
is a numerical artifact caused by the necessity of solving the equations in a finite
radial domain. 
	\label{fig:DispersionFunction}}
\end{figure*}

Figure~\ref{fig:DispersionFunction} illustrates the numerically computed dispersion
function and its variation with the spectral parameters $k$ and $\omega$ for the
fundamental azimuthal order $m=1$. The frequency and wavenumber dependence are
characterized with a dimensionless frequency $\omega \, r_0 V_0^{-1}$ and a
dimensionless wavenumber $kr_0$. A smooth gradient in frequency has been removed
from the image for purely illustrative purposes. This has been accomplished by
multiplying the reciprocal of the dispersion function by a pure function of frequency,
$f(\omega) = \exp\left(-0.2 \; \omega \, r_0 V_0^{-1}\right)$. One can clearly
recognize the presence of strong fast-mode resonances which appear as bright
``ridges" of high power. As we will soon see, each ridge corresponds to
resonant fast waves with differing number of peaks in the modulus of their
eigenfunctions as a function of radius. The lowest-frequency ridge has one peak
in its eigenfunction and we designate its radial order as $n=0$. Each ridge of
sequentially higher frequency has one additional peak and a corresponding radial
order of one higher, i.e., $n=1$, 2, 3, and so forth. Traditionally, the radial
order of a mode (or ridge) specifies the number of radial nodes in its eigenfunction.
However, because of resonant absorption, our eigenfunctions are complex and lack
nodes where both the real and imaginary part vanish. Therefore, we use the number
of maxima in the modulus of the eigenfunction instead. The radial order $n$ is
defined to be one less than the number of peaks.

These radial resonances arise from waves trapped between two radial turning points.
The outer turning point is caused by refraction from the increasing Alfv\'en speed
with height. The inner turning point is due to the combined effect of reflection
from the Alfv\'en singularity and from the geometrical increase in the azimuthal
wavenumber $m/r$ for small radii. For a given radial order, the frequency approaches
a nonzero value as the axial wavenumber approaches zero. This is due to the asymptotic
form of our Alfv\'en speed profile for large radius. Refraction requires oblique
propagation and waves that propagate purely radially $k=0$ are unrefracted and
untrapped. Hence as the axial wavenumber  $k$ approaches zero, the outer turning
point approaches infinity and the radial wavelength becomes infinity large.
Since the cavity extends to large radii, the Alfv\'en speed is nearly a linear
function throughout most of the wave cavity, $\va \approx V_0 \, r/r_0$. Using
this asymptotic form, equation~\eqnref{eqn:WKB_form} reveals that the wave's frequency
must approach a finite value as $k$ vanishes, $\omega \to m V_0 r_0^{-1}$.

\begin{figure*}
	\epsscale{1.0}
	\plotone{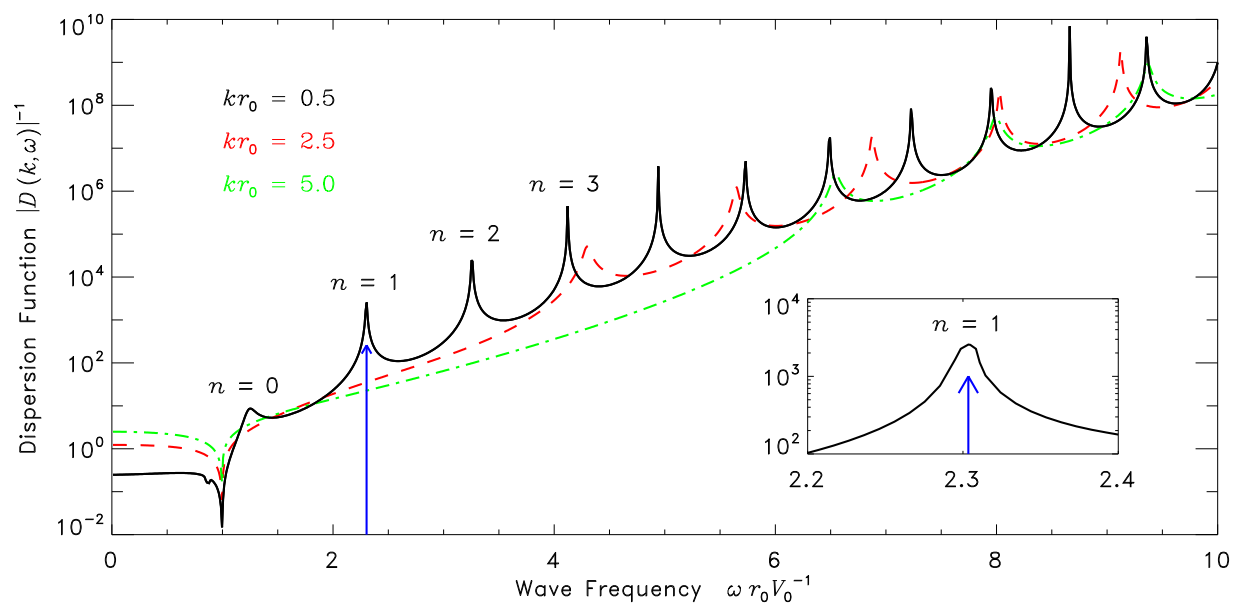}
	\caption{\small Three cuts at constant wavenumber through the reciprocal
of the modulus of the dispersion function shown in Figure~\ref{fig:DispersionFunction}.
The black solid curve illlustrates the dispersion function for a low wavenumber,
$kr_0 = 0.5$ and the dashed red and dot-dashed green curves show successively higher
wavenumbers, $kr_0 = 2.5$ and $5.0$ respectively. The $n=1$ mode, indicated with
the blue arrow, is shown in a zoom-in view in the inset.
	\label{fig:DispersionCuts}}
\end{figure*}

All of the modes are damped due to resonant absorption by the Alfv\'en waves and
hence have complex eigenfrequencies or wavenumbers. Thus, for real frequency and
real wavenumber (as shown in the figure), the dispersion function is never truly
zero. Instead, as the frequency and wavenumber pass near a complex zero in the
dispersion function, the power peaks with a Lorentzian profile. The width of the
profile is a direct measure of the damping rate. Figure~\ref{fig:DispersionCuts}
shows cuts through the reciprocal of the dispersion function at three representative
values of the axial wavenumber. Each ridge appears as a large amplitude Lorentzian
spike. This profile shape is revealed in the inset where the mode marked with the
blue arrow is shown over a narrow range of frequency.

The lowest radial-order ridge in the power spectrum illustrated in
Figure~\ref{fig:DispersionFunction} spans only a short range of wavenumbers before
it fades into the background. This ridge appears near a nondimensional frequency
of $\omega r_0 V_0^{-1} = m = 1$ in the lower-left corner of the power spectra.
This short feature of high power corresponds to the fundamental radial order $n=0$.
As we will see, this mode is strongly damped by resonant absorption and at higher
wavenumbers ($kr_0 > 1$), the ridge widens so much in spectral space that it becomes
indistinguishable from the background. When we examine eigenfunctions in detail
in the next section, we will find that the energy density of the fundamental radial
mode has a single extremum located at the Alfv\'en singularity. Each ridge of
successively higher frequency and radial order ($n=1$, $n=2$, $n=3$, and so forth)
adds an additional peak to the energy density as a function of radius. These radial
overtones are only weakly damped and thus have narrow ridges and the power varies
by orders of magnitude between the inter-ridge background and the ridge peaks
(see Figure~\ref{fig:DispersionCuts}).

Due to the fact that we solve the ODEs by Runge-Kutta integration, we expect that
numerical inaccuracies should arise when we must integrate very long distances.
The region of spectral space near the low-wavenumber base of the lowest-frequency
ridge is a region where such long integrations become necessary. This numerical
difficulty arises for two compounding reasons. As the dimensionless frequency
approaches $m$ the resonant field line moves infinitely far away in radius.
Similarly, the outer refraction point or turning point, $r_2$, of the fast wave
moves farther and farther away from the axis as the wavenumber becomes small,
$r_2 \approx r_0 V_0^{-1} \, \omega/k$. For both reasons, as the axial wavenumber
vanishes and the dimensionless frequency approaches $m$, we must move the outer
boundary of our computational domain, $r_{\rm out}$, further and further away.
This means that our numerical integrations must cross longer and longer distances
and as numerical errors accumulate, these long distance integrations become
increasingly inaccurate.  In the very small spectral region at the base of the
$n=0$ ridge where this inaccuracy becomes unacceptable, we saturate the image
with pure white.


\subsection{Eigenwavenumbers}
\label{subsec:eigenwavenumbers}

\begin{figure*}
	\epsscale{1.0}
	\plotone{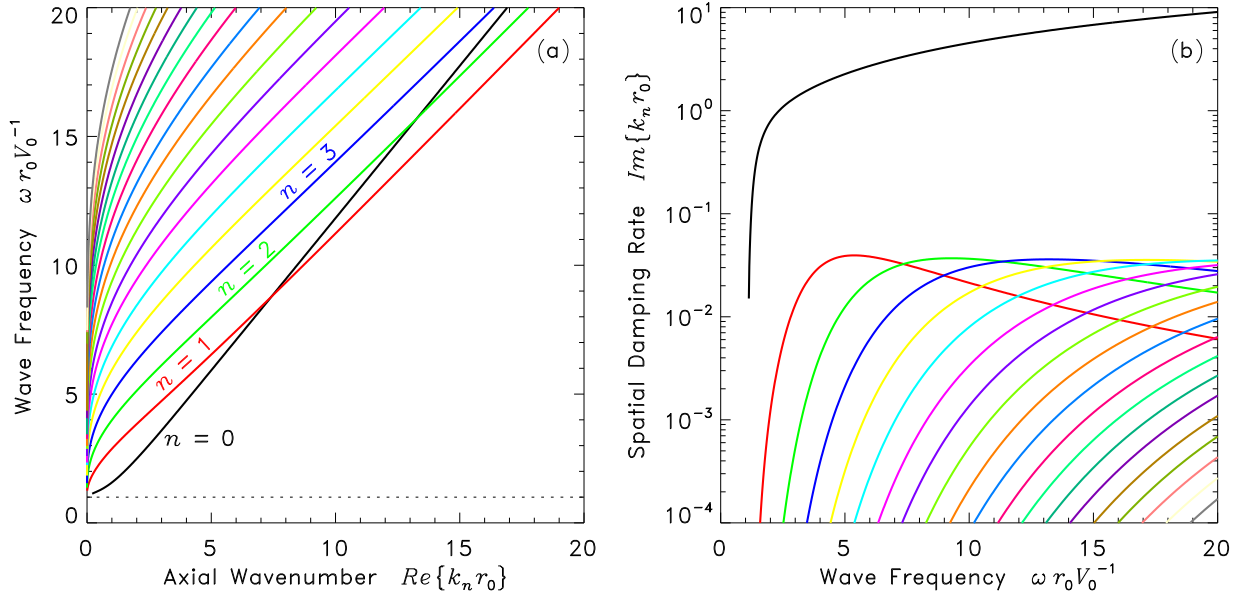}
	\caption{\small (a) Real and (b) imaginary parts of the eigenwavenumber
displayed as a function of frequency for $m=1$. Each color curve corresponds to
a different radial order, with black indicating the fundamental radial order,
$n=0$. Higher radial orders are illustrated with red ($n=1$), green ($n=2$),
blue ($n=3$), and so on. The fundamental radial order has a different asymptotic
slope for large wavenumber than the higher radial orders. Further, the spatial
damping rate for the fundamental radial order is much larger. Note, even though
the dispersion relation (panel a) for the fundamental mode crosses those of higher
order, this is {\bf not} an avoided crossing because the eigenwavenumbers are
complex and the modes have distinct imaginary parts. The dotted horizontal line
indicates the dimensionless frequency equals $m$.
	\label{fig:DispersionRelation}}
\end{figure*}

We find the complex eigenwavenumbers using a complex root finder that utilizes
a numerical routine which calculates the dispersion function for any requested
frequency and wavenumber. At a given frequency, the root finder starts from an
initial guess for the eigenwavenumber and iterates until it converges to a complex
zero in the dispersion function. Figure~\ref{fig:DispersionRelation} shows the
real and imaginary parts of the resulting axial wavenumber as a function of
position along each ridge. The black curve corresponds to the fundamental radial
mode ($n=0$) and the sequence of colored curves is a sequence of radial order
$n$ (red: $n=1$, green: $n=2$, blue: $n=3$, etc.). Figure~\ref{fig:DispersionRelation}$a$
shows the dispersion relation, i.e., the frequency as a function of the real part
of the wavenumber, while Figure~\ref{fig:DispersionRelation}$b$ shows the spatial
damping rate (the imaginary part of the wavenumber) as a function of frequency.
Since, the eigenwavenumber, $k_n(\omega)$, only appears as its square $k_n^2$ in
the ODE~\eqnref{eqn:PhiODE}, for each eigenvalue there are a pair of solutions
$\pm \sqrt{k_n^2}$, one corresponding to forward-propagating waves (i.e., phase
moves in the positive $y$-direction) and the other to backward propagating waves.
Since the two solutions have the same radial eigenfunction, we only show the
forward-propagating wave with a positive real part of its eigenwavenumber.

The radial overtones ($n>0$) all have dispersion relations and spatial damping
rates with similar behavior. As the frequency increases, the real part of the
wavenumber increases, initially quite quickly, and then approaches a slope common
to all overtones. The spatial damping rate of all overtones is quite small and
peaks at a frequency that depends on the radial order. Interesting, the dimensionless
damping rate for all radial orders peaks at roughly the same value, $\sim 4\times 10^{-2}$.
The radial fundamental mode ($n=0$) behaves discordantly. The dispersion relation
is convex instead of concave and asymptotically approaches a steeper slope than
the overtones. More importantly, the fundamental mode is heavily damped. Resonant
absorption produces a spatial decay rate that is orders of magnitude larger than
that evinced by the overtones.

\begin{figure*}
	\epsscale{1.0}
	\plotone{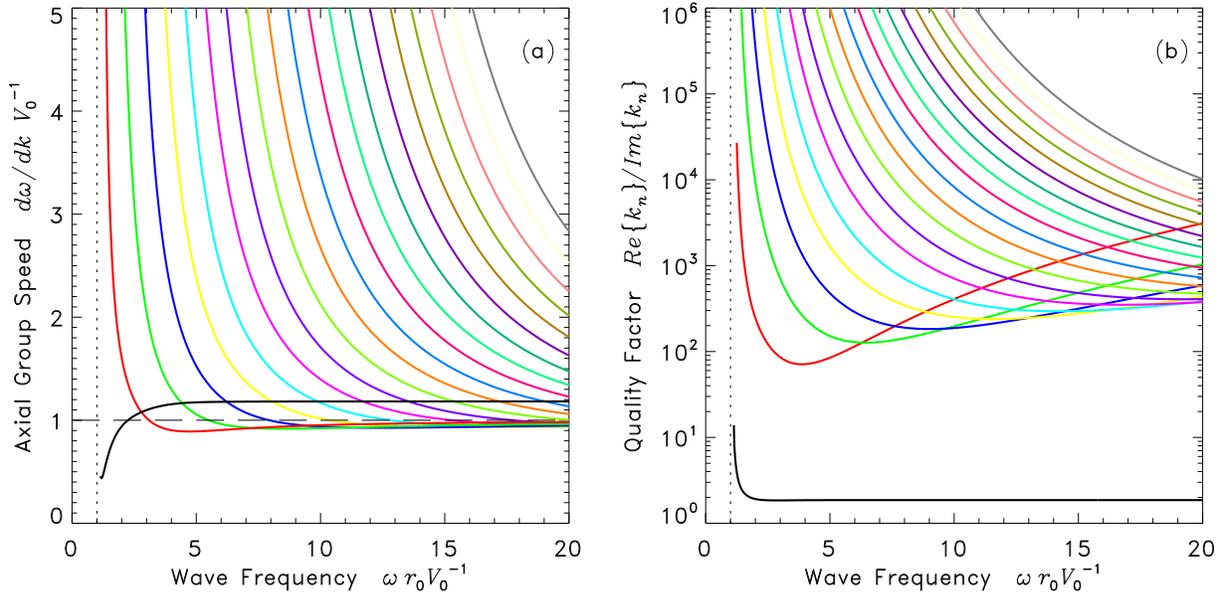}
	\caption{\small (a) Axial group speed and (b) quality factor for the eigenmodes
as a function of dimensionless frequency and radial order. The azimuthal order is
unity, $m=1$. The colors have the same meaning as in Figure~\ref{fig:DispersionRelation}.
The group velocity is defined as the slope of the curves in Figure~\ref{fig:DispersionRelation}a.
Asymptotically for high frequency, the group speed approaches a constant. For the
radial overtones ($n>0$), this asymptotic value is equal to the Alfv\'en speed at
the axis, $V_0$. For the fundamental mode, the asymptotic value is somewhat higher.
The quality factor, $Q$, is the ratio of the real to imaginary parts of the eigenwavenumber
and signifies the number of wavelengths the wave travels before it is significantly
diminished in amplitude. The fundamental mode is heavily damped and travels only
a few wavelengths before being absorbed ($Q \sim 2$). The radial overtones are all
weakly damped ($Q > 100$) and the quality factor achieves a minimum at a frequency
that increases as the order increases.
	\label{fig:GroupSpeed}}
\end{figure*}

Figure~\ref{fig:GroupSpeed}a shows the axial group speed, $v_{\rm grp}$, of each
radial order as a function of frequency,

\begin{equation}
	\frac{1}{v_{\rm grp}} \equiv Re\left\{\dpar[k_n]{\omega}\right\} \; .
\end{equation}

\noindent The overtones ($n>0$) behave as one would expect. Waves with low axial
wavenumber propagate outward nearly radially and refract back toward the axis at
a turning point located at large radius. Such waves sample regions with high
Alfv\'en speed and thus have a high axial group speed. As the frequency and wavenumber
increase, the turning point moves inward, confining the mode to the region of low
Alfv\'en speed near the axis. Hence, the group speed asymptotically approaches a
value of $V_0$ as the frequency increases. The radial fundamental modes defies this
trend. Its group speed increases with frequency and approaches a constant value that is
larger than $V_0$.

The magnitude of the spatial damping rate can be put into context by considering
the quality factor $Q$ which is defined as the ratio of the real and imaginary part
of the eigenvalue,

\begin{equation}
	Q \equiv \frac{Re\left\{k_n\right\}}{Im\left\{k_n\right\}} \; .
\end{equation}
 
\noindent The quality factor is a measure of the number of wavelengths that the 
wave travels before it suffers significant attenuation. The quality factor for
each radial order is shown in Figure~\ref{fig:GroupSpeed}b. The overtones are all
weakly damped and travel hundreds of wavelengths before significant amplitude
decay occurs. As such, the quality factor for these modes exceeds 100 for all
frequencies. On the other hand, the radial fundamental mode has a quality factor
of nearly 2 at almost all frequencies. This wave only travels a single wavelength
before losing most of its energy through resonant coupling with the Alfv\'en
waves.


\subsection{Eigenfunctions}
\label{subsec:eigenfunctions}

The eigenfunctions can be generated by evaluating the outer solution at the
eigenwavenumber,

\begin{equation}
	\Phi_{mn}(r \vsep \omega) = \Phi_\infty(r \vsep k_n(\omega), \omega) \; .
\end{equation}

\begin{figure*}
	\epsscale{1.0}
	\plotone{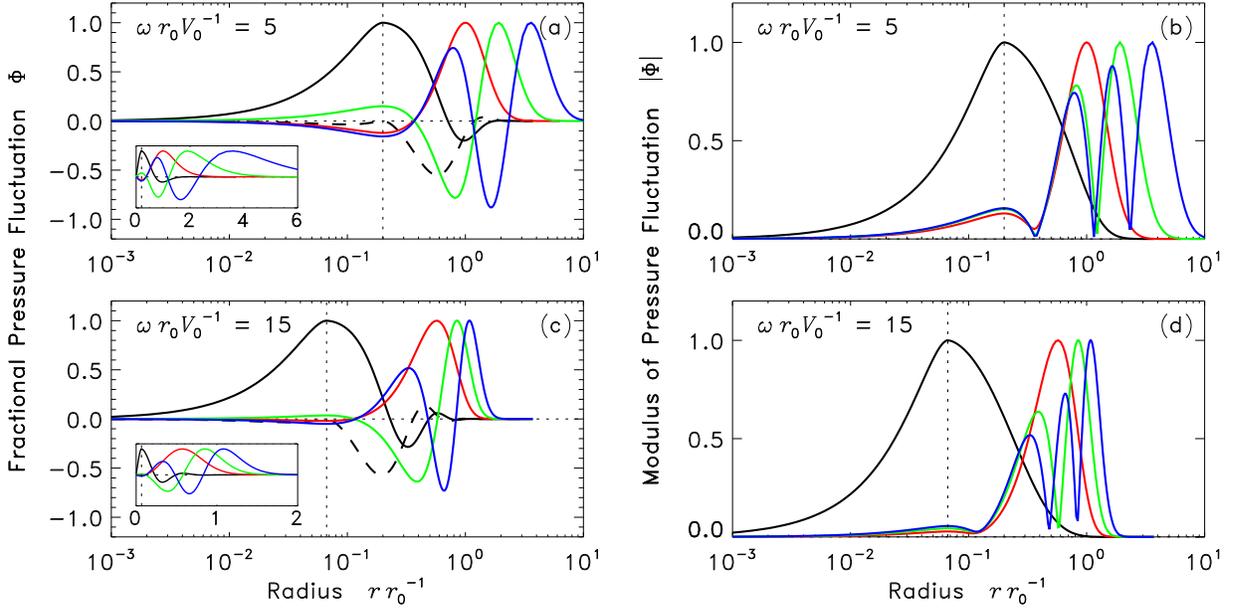}
	\caption{\small The fractional magnetic-pressure fluctuation,
$\Phi_{mn}(r \vsep \omega)$, as a function of radius, for the eigenfunctions of the
first four radial orders and for $m=1$. The different colors of
the curves indicate the radial order (black: $n=0$, red: $n=1$, green: $n=2$, and
blue: $n=3$). In all panels the vertical dotted line indicates the radius at which
the Alfv\'en singularity occurs for the given frequency. (a) The pressure eigenfunctions
for a dimensionless frequency of $\omega r_0 V_0^{-1} = 5$. The solid curves display
the real part of the complex eigenfunction and the dashed-curve shows the imaginary
part for the radial fundamental ($n=0$). The imaginary parts of the other orders
have been omitted since their eigenfunctions are nearly real due to weak damping.
The inset shows the real part of the eigenfunctions on a linear radius scale.
(b) The modulus of the complex eigenfunctions. From this figure one can easily recognize
that the radial order indicates the number of peaks appearing in the modulus (e.g., $n=0$
has 1 peak, $n=1$ has 2 peaks and so forth).
(c) Pressure fluctuation for eigenfunctions with a dimensionless frequency of
$\omega r_0 V_0^{-1} = 15$. The eigenfunctions all have the same structure as those
at lower frequency, but are compressed towards the origin.
(d) The modulus of the eigenfunctions for the higher frequency.
	\label{fig:EigFunc_Phi}}
\end{figure*}

\noindent Figures~\ref{fig:EigFunc_Phi}a and c illustrate the fractional magnetic-pressure
fluctuation for eigenfunctions with radial orders $n \leq 3$ for two different
frequencies. The color of the curves indicates the eigenmode's radial order (see
the caption). As the radial order increases, the axial wavelength of the mode
increases (see Figure~\ref{fig:DispersionRelation}) and the wave refracts at a
larger radius where the Alfv\'en speed is larger. Thus, the higher-order overtones
have eigenfunctions with larger wave cavities that extend higher in radius. To
capture this increase in size of the wave cavity in a common figure, we have plotted
the eigenfunctions using a logarithmically scaled radius axis. A linear scaling
is also provided in a small inset. All eigenfunctions are complex quantities
because of the damping caused by resonant absorption. However, the damping is
exceedingly weak for all modes other than the radial fundamental mode. Thus, only the
$n=0$ mode has a significant imaginary component. The real part of each eigenfunction
is displayed using solid curves and the imaginary part of the fundamental is
shown with a dashed curve. Since the eigenfunctions are complex, it is not
obvious that the radial order indicates the number of nodes in the eigenfunction.
However, if we examine the modulus of the eigenfunctions, as illustrated in
figures~\ref{fig:EigFunc_Phi}b and d, the trend becomes clear. The modulus of
each complex eigenfunction has a number of peaks in radius equal to $n+1$. Thus,
even though the radial fundamental mode has an eigenfunction with significant real
and imaginary parts, and those parts may separately have nodes, the modulus
is singly lobed with one extrema.

From the power series expansions of the two solutions valid at the singularity
(see Appendix~\ref{subsec:interior}), one can easily determine that the pressure
fluctuation, $\Phi$, remains finite and has vanishing derivative. The regular
solution and its derivative both vanish, so the solution is dominated by the
irregular solution

\begin{eqnarray} \label{eqn:Phi_sing}
	\Phi \, \propto \, \Phi_{\rm irr} &=& 1 + \frac{k^2 r_0^2}{2} \, (r-\ra)^2 \ln(r-\ra) + \cdots
		   \to 1 \; ,
\\ 		\label{eqn:dPhi_sing}
	\der[\Phi]{r} \, \propto \, \der[\Phi_{\rm irr}]{r} &=& k^2 r_0^2 \, (r-\ra) \ln(r-\ra) + \cdots
		   \to 0 \; .
\end{eqnarray}

\noindent From these two expansions, we immediately deduce that the Alfv\'en
radius must be a local peak in the eigenfunction.  This prediction is born out
in an examination of the numerically calculated eigenfunctions presented in
Figure~\ref{fig:EigFunc_Phi}. The influence of the Alfv\'en singularity
essentially adds an additional peak to all of the eigenfunctions of the fast
modes. This suggests that the $n=0$ mode (with only a single peak at the Alfv\'en
singularity) only exists because of the presence of the singularity. The radial
fundamental mode has a different physical origin than the higher radial orders.

While the magnetic-pressure fluctuation remains finite at the singularity, the
velocity components of the eigenfunction do not. From the leading-order behavior
at the singularity, equation~\eqnref{eqn:u_ODE} indicates that both
the axial and radial velocity are singular,

\begin{eqnarray}
	u_y = \frac{\va^2}{\omega^2-\oma^2} \; ik \Phi &\sim& \frac{1}{r-\ra} + \cdots \; ,
\\
 	u_r = \frac{\va^2}{\omega^2-\oma^2} \; \der[\Phi]{r} &\sim& k^2 r_0^2 \, \ln(r-\ra) + \cdots \; .
\end{eqnarray}

\begin{figure*}
	\epsscale{1.0}
	\plotone{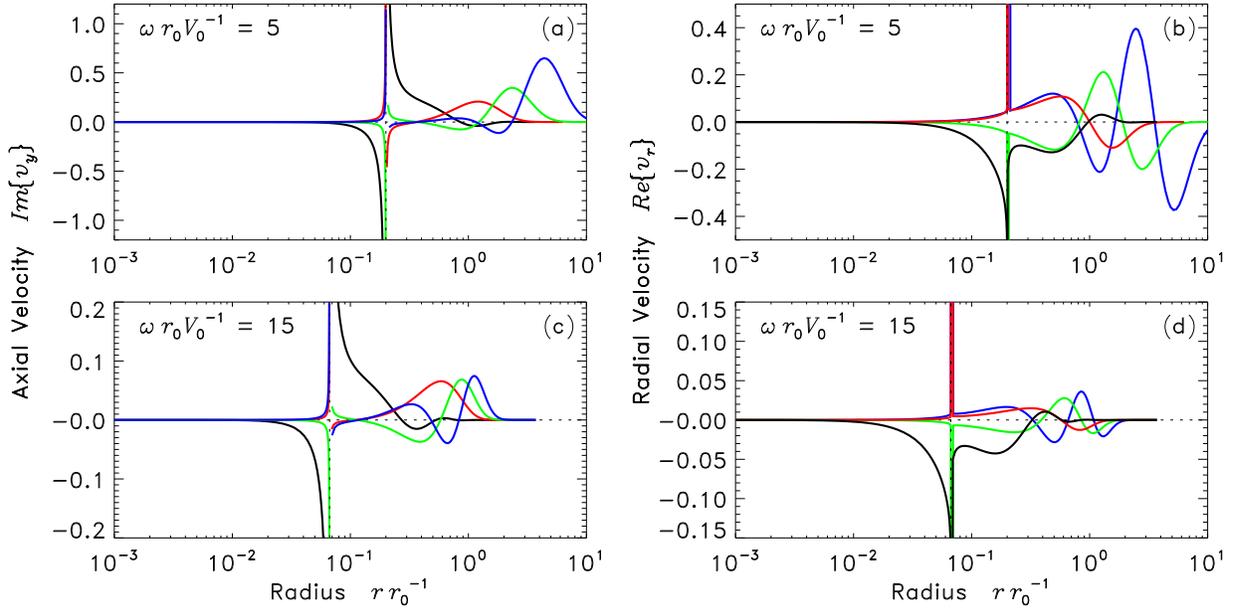}
	\caption{\small The velocity eigenfunctions for the same modes illustrated
in Figure~\ref{fig:EigFunc_Phi}. The colors have the same meaning as in that figure.
The left-hand panels (a) and (c) display the imaginary part of the axial velocity
component and the right-hand panels (b) and (d) present the real part of the radial
velocity component.  The upper panels show a small frequency and the lower panels
show a large frequency. Both velocity components are complex quantities, but for
the radial overtones (which have weak damping), the axial component is almost purely
imaginary and the radial component is nearly real. All velocity components are singular
at the Alfv\'en radius (which is indicated by the vertical dotted line). The axial
velocity component has a hyperbolic divergence, $\sim 1/(r-\ra)$ while the radial
component has a logarithmic divergence $\sim \log(r-\ra)$.
	\label{fig:EigFunc_v}}
\end{figure*}

\noindent Figure~\ref{fig:EigFunc_v} illustrates both of these velocity components.
When the fractional magnetic-pressure fluctuation, $\Phi$, is purely real, the axial
velocity is purely imaginary and the radial velocity is purely real.  Hence, in
Figure~\ref{fig:EigFunc_v} we display the imaginary part of the axial velocity and
real part of the radial velocity.


\section{Discussion}
\label{sec:Discussion}

The dispersion relation and spatial damping rate (see Figures~\ref{fig:GroupSpeed}a,b)
clearly demonstrate that radial fundamental mode ($n=0$) has a very different physical
nature than all of the higher order radial modes. We reiterate that the radial
fundamental only exists because of the singularity. If we were to slowly perturb
the Alfv\'en speed profile to one where the singularity vanished (for example
$\va \propto r$), the radial fundamental mode would also disappear, while the overtones
would remain.


\subsection{Wave Cavities}
\label{subsec:cavities}

\begin{figure*}
	\epsscale{1.0}
	\plotone{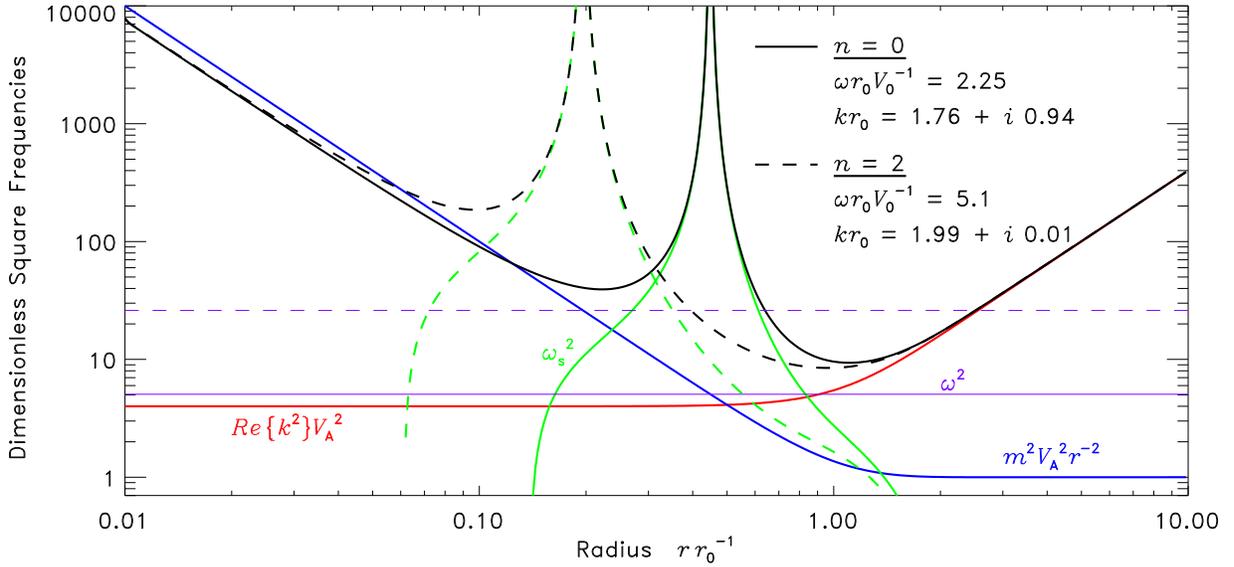}
	\caption{\small A propagation diagram illustrating that the fundamental
radial mode is a fast surface wave, whereas the radial overtones are fast body waves.
Two particular modes with $m=1$ are presented. The solid curves are those for a $n=0$ mode with
a dimensionless frequency of 2.55 and a complex wavenumber of $kr_0 = 1.76 + i \, 0.94$.
The dashed curves are for a $n=2$ mode with a frequency of 5.1 and wavenumber of
$kr_0 = 1.99 + i \, 0.01$.  Both modes have $Re\left\{k^2\right\} = 4.0$. Thus the
solid red and blue curves show common values for the frequencies $k\va$ and $m r^{-1} \va$. The two
horizontal violet curves indicate the two mode frequencies. For the $n=2$ mode, it
is clear that there exists a range of radii for which the frequency exceeds the
critical frequency and there exists a cavity where the wave propagates. The overtone
is therefore primarily a body wave.  The fundamental radial mode ($n=0$) has a frequency that
nowhere exceeds the critical frequency. Hence is must be a surface wave that exists
on the only discontinuity present, the Alfv\'en singularity.
	\label{fig:Surface_Wave}}
\end{figure*}

The disparate behavior between the fundamental radial mode and the radial overtones
is primary due to the fact that the overtones reside in a fast wave cavity whereas
the fundamental is a surface wave that resides on the singularity.  All of the overtones
($n>0$) have frequencies that are sufficiently high that they possess a cavity where
the wave is propagating as indicated in Figure~\ref{fig:Propagation_Diagram}. Thus,
the overtones are all primarily fast body waves. Further, the Alfv\'en singularity
lies just inside the inner turning point and thus lies outside of the fast wave cavity.
This explains why the the resonant coupling between the overtones and the Alfv\'en
waves is inefficient and the overtones are weakly damped. The singularity lies in
the evanescent tail of the overtones where the mode's amplitude is relatively
low. The radial fundamental $n=0$ on the other hand is nowhere propagating and lacks
a fast wave cavity. Figure~\ref{fig:Surface_Wave} provides a propagation diagram for
two modes $n=0$ and $n=2$.  Both have eigenvalues $k^2$ with identical real parts,
$Re\left\{k^2\right\}=4.0$. Of course, they have different frequencies and different
imaginary parts of their eigenvalues. Nowhere does the fundamental have a frequency
that exceeds the critical frequency. Thus, this mode is a surface wave that lives
on the only available discontinuity, the Alfv\'en singularity. Since this mode's
energy density is concentrated at the singularity, resonant absorption is efficient.


\subsection{Restoring Forces}
\label{subsec:restoring_forces}

\begin{figure*}
	\epsscale{1.0}
	\plotone{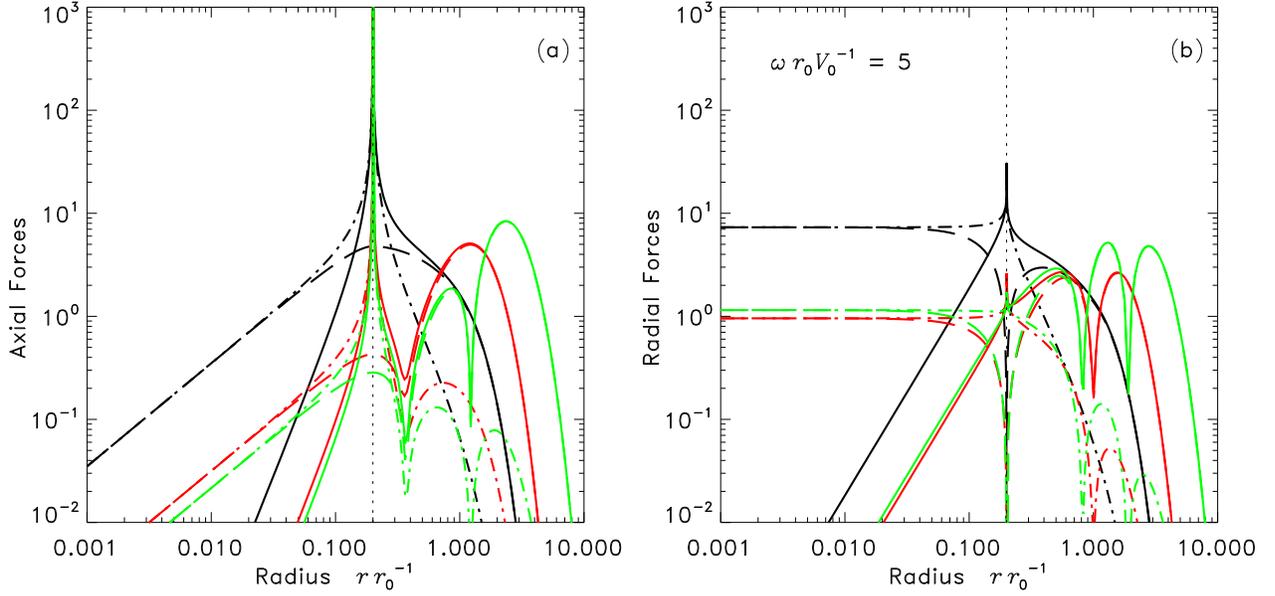}
	\caption{\small The magnitudes of the two restoring forces, magnetic pressure
and magnetic tension, in (a) the axial direction and (b) the radial direction. The
black curves indicate the forces for the radial fundamental mode $n=0$, while the
red and green curves indicate the first two overtones, $n=1$ and $n=2$ respectively.
All forces are for modes with the same nondimensional frequency, $\omega r_0 V_0^{-1} = 5$,
and azimuthal order, $m=1$. The modulus of the magnetic pressure force is illustrated
with the dashed curves and the modulus of the tension with the dot-dashed curves.
The modulus of the total restoring force (pressure plus tension) is shown with solid
curves. Clearly, the magnetic pressure dominates over most of the wave cavity. The
only exceptions are at the Alfv\'en singularity (indicated with the vertical dotted
line) and near the origin. Near the singularity, the magnetic tension dominates. Near
the origin the two forces have the same magnitude but are almost 180 degrees out of
phase such that they nearly cancel when summed.
	\label{fig:Forces}}
\end{figure*}

Fast waves have two potential restoring forces that usually act in concert, magnetic
pressure and magnetic tension. Figure~\ref{fig:Forces} illustrates the magnitudes of
these two forces for three modes of different radial order but the same frequency.
A careful examination of the restoring forces in both the axial and radial directions
reveals that the magnetic pressure dominates everywhere except at the Alfv\'en singularity
and near the origin. At the singularity magnetic tension becomes the most important
force. Near the origin, the tension and pressure are of equal magnitude but nearly
180 degrees out of phase such that they nearly cancel each other. Since the overtones
have radial cavities that extend well-above the Alfv\'en singularity, over most of
the region where the mode has significant amplitude, the magnetic pressure is the
primary restoring force. Hence, the overtones are primarily fast pressure waves
with weak contributions from the tension. The radial fundamental mode on the other
hand has an eigenfunction that is confined to the immediate vicinity of the singularity.
Thus, not only is the fundamental primarily a fast tension wave with weak pressure
contributions, its suffers extensive resonant absorption because it is confined to
a region of strong coupling.


\subsection{Observational Implications}
\label{subsec:observations}

Our assumption that the axial direction is infinite and ignorable is not trivial.
The primary effect is that all axial wavenumbers are allowed because of the infinite
domain. This allows us to consider the axial wavenumber as a ``radial eigenvalue" and
the frequency as a free-parameter. On the other hand, if there were boundaries placed
along the axial coordinate, the wavenumber would be quantitized and fixed in order
to satisfy the axial boundary conditions. This would require
that the frequency represent the radial eigenvalue.

For most of the fast modes this is a serious constraint. They have such low damping
rate arising from resonant absorption that they can travel very far before having
their signal attenuated. For example, if we consider a model arcade with a typical
length scale of $r_0 = 100$ Mm, the smallest spatial decay {\bf length} for the
overtones is roughly $r_0 / (4 \times 10^{-2}) = 2500$ Mm. Assuming a typical arcade
whose length is on the order of a solar radius, these waves would sense both boundaries.
Thus, they would bounce back and forth multiple times before being dissipated. The
low-amplitude ``decayless oscillations" may be the result of such fast wave oscillations.
One or more excitation events generate a set of radial overtones that interfere
with each other and with themselves as they criss-cross the axial cavity.

The radial fundamental mode is different. This mode only travels a few wavelengths before
significant decay occurs. Thus, this mode would die before reaching the edges of a
long arcade and its axial wavenumber would not be quantitized to an obvious degree.
Once excited these waves would dissipate rather quickly into Alfv\'en waves and would
never travel far from where they were generated. Thus, the radial fundamental mode can
only be observed as coronal loop oscillations if those oscillations are excited locally.

Loop oscillations have to date only been observed by their motion on the plane of
the sky. Thus, what is observed is a linear combination of the two transverse velocity
components, and observed polarization depends on the angle between the line of sight
and the arcade's axis. The eigenfunction for either velocity component is singular
at the Alfv\'en radius. Note, however, this singularity should never be observed,
as a pure eigenmode is never excited. Instead, all physically realizable wave fields
are composed of a linear combination of eigenfunctions represented by an integral
over frequency. Each frequency component in the integration has a singularity at
a different radius. So, even though the integrand has singularities the integral
must remain finite at all radii. For wave packets comprised of a narrow band of
frequencies we expect a large, but finite, radial response at the Alfv\'en radius
concomitant with the central frequency of the band. But, the wave packet may also
have a significant response at higher radii corresponding to peaks in the amplitude
of the velocity eigenfunctions if the packet possesses radial overtones in its makeup.
The motion of a single coronal loop, however, will only reveal the motion at a single
shell or magnetic surface. It could very well be that the largest field line displacements
remain invisible if the radius of the observed loop does not coincide with the Alfv\'en
radius for the dominant frequency of the excited waves.


\acknowledgements

This work was supported by NASA through grants NNX14AG05G, NNX14AC05G, and NNX17AM01G.
RJ would like to acknowledge the support of the University of Sheffield (UK). We
thank Bengt Fornberg for a helpful discussion concerning analytic continuation of
numerical solutions.



\appendix

\section{Solutions near the Singular Points}
\label{sec:Singularities}

The governing ODE~\eqnref{eqn:PhiODE} has three (or more) singular points in radius:
the origin ($r=0$), infinity ($r \to \infty$), and one or more Alfv\'en resonances
($r = \ra$). For monotonic Alfv\'en frequency profiles, like that which arises from
equation~\eqnref{eqn:VaProfile} and illustrated in Figure~\ref{fig:AlfvenProfile},
there is at most a single Alfv\'en radius for any particular frequency. The solutions
that are regular at the origin and at infinity can be obtained through standard
expansions and asymptotic analyses. As such, they will be discussed here only briefly.
The interior singularity at the Alfv\'en radius is more unusual and warrants a more
detailed examination.


\subsection{Solution near the Origin}
\label{subsec:origin}

A power-series solution to equation~\eqnref{eqn:PhiODE} that is valid near the
origin can be found by expanding the square of the local wavenumber, $K^2$, in
an expansion for small radius, $r << \inf\left(r_0, \, k^{-1}\right)$, and using
the method of Frobenius to build the series solution term by term. To lowest
order, the two solutions to equation~\eqnref{eqn:PhiODE} behave like $r^{\pm m}$.
We present the first few terms in the power series for the regular, well-behaved
solution,

\begin{eqnarray}
	\Phi_0 \left(r \vsep k,\omega\right) &=& (r/r_0)^m \left[1 + a_2 (r/r_0)^2 + a_4 (r/r_0)^4 + \cdots\right] \;
\\
	a_2 &=& -\frac{\Omega^2 - k^2 r_0^2 + 2m\alpha}{4(m+1)}  \;
\\
	a_4 &=& -\frac{\Omega^2 - k^2 r_0^2 + 2(m+2)\alpha}{8(m+2)} \; a_2
			+ 2m\alpha \; \frac{(m+4)q_2 - 2\alpha}{16(m+2)}  \; ,
\\
	\Omega &\equiv& \frac{\omega r_0}{V_0} \; , \;\;\;\; \alpha \equiv \frac{\Omega^2}{m^2} \; , \;\;\;\; 
		q_2 \equiv \frac{r_0^2}{2! V_0^2} \left. \dern[V_A^2]{r}{2} \right|_{r=0} \; .
\end{eqnarray}


\subsection{Asymptotic Solution for Large Radius}
\label{subsec:infinity}

In the limit of large radius, $r >> \sup\left(r_0, \, k^{-1}\right)$, the outer
solution $\Phi_\infty$ behaves like a modified Bessel function. This can be verified
by recognizing that $\va \to V_0 (r/r_0)$ and the ODE~\eqnref{eqn:PhiODE} becomes
Bessel's equation \citep{Hindman:2015},

\begin{equation}
	\dern[\Phi_\infty]{r}{2} + \frac{1}{r} \der[\Phi_\infty]{r} - \left(\frac{\nu^2}{r^2} + k^2\right) \Phi_\infty = 0 \; ,
\end{equation}

\noindent where we define the constant $\nu^2 \equiv m^2 - \Omega^2$. The solution
that is well-behaved for large radius is the $K$ function (modified Bessel function
of the second kind),

\begin{equation}
	\Phi_\infty(r \vsep k,\omega) \to K_\nu(\pm k r) \; ,
\end{equation}

\noindent where the sign of the argument is chosen to ensure that the real part is
positive (otherwise the solution diverges exponentially for large $r$).

In order to use the asymptotic solution to initialize the numerical integrations
in our analysis (and hence satisfy the radial boundary condition at infinity), we
need to be able to evaluate the $K$ Bessel function for complex frequencies and
wavenumbers. Complex frequencies impose complex azimuthal orders, $\nu$, and complex
wavenumbers impose complex arguments.  Most standard Bessel function algorithms
are incapable of dealing with these complications. Hence, we use an asymptotic
expansion that is valid for large values of the argument,

\begin{equation}
	\Phi_\infty(r \vsep k,\omega) = \sqrt{\frac{\pi}{2kr}} \; e^{-kr} \; \sum_{n=0}^{N-1} \frac{b_n}{n!(8kr)^n} \; ,
\end{equation}

\noindent with coefficients given by

\begin{eqnarray}
	b_0 &\equiv& 1 \; ,
\\
	b_n &\equiv& \left[4\nu^2 - (2n-1)^2\right] \, b_{n-1} \; \; \; {\rm for} \; \;  n > 0 \; .
\end{eqnarray}

\noindent The number of terms $N$ is chosen to optimize the accuracy of the expansion
and the radius at which to initiate the integration is chosen to ensure a minimal
degree of accuracy.


\subsection{Solution near the Alfv\'en Singularity}
\label{subsec:interior}

The Alfv\'en resonances ($r = \ra$) are interior singularities that are logarithmic
regular singular points \citep[e.g.,][]{Uberoi:1972, Appert:1974, Goedbloed:1971},
much like the origin in Bessel's equation. The two solutions are therefore a regular,
well-behaved solution $\Phi_{\rm reg}$ and an irregular solution $\Phi_{\rm irr}$
with a logarithmic singularity. Both will be needed. Through a Frobenius expansion
around the singularity, one can obtain expansion coefficients. The series are written
in terms of a nondimensional local coordinate $\varepsilon = (r - \ra)/r_0$. The
first handful of terms are as follows

\begin{eqnarray}
	\Phi_{\rm reg}(r \vsep k,\omega) &=& \varepsilon^2 + \frac{2}{3} \, A_0 \, \varepsilon^3 +
		\frac{1}{4} \left(A_0^2 - A_1 -\frac{k^2 r_0^2}{2}\right) \, \varepsilon^4 + O\left(\varepsilon^5\right) \; ,
\\
	\Phi_{\rm irr}(r \vsep k,\omega) &=&  \frac{k^2 r_0^2}{2} \, \Phi_{\rm reg} \, \ln\left(\varepsilon\right) + 
		1 - \frac{1}{6}\left(\frac{5}{6} A_0 + \Lambda^\prime(0)\right) \varepsilon^3 + O\left(\varepsilon^4\right) \; ,
\end{eqnarray}

\noindent where the primes denote differentiation with respect to $\varepsilon$
and the following definitions have been made,

\begin{eqnarray}
	\Lambda(\varepsilon) &\equiv& r_0^2 \, K^2 =  r_0^2 \; \frac{\omega^2-\omega_A^2}{\va^2} \; ,
\\
	A_0 &\equiv& \frac{\Lambda^{\prime\prime}(0)}{2 \Lambda^\prime(0)} + \frac{r_0}{\ra} \; ,
\\
	A_1 &\equiv& \frac{\Lambda^{\prime\prime\prime}(0)}{3 \Lambda^\prime(0)}
		- \left(\frac{\Lambda^{\prime\prime}(0)}{2 \Lambda^\prime(0)}\right)^2
		- \left(\frac{r_0}{\ra}\right)^2 \; .
\end{eqnarray}

\noindent In developing these series expansions, we have assumed that the derivative
of the local Alfv\'en frequency is nonzero at the singularity such that to
lowest order $\omega^2 - \omega_A^2 \sim \varepsilon$ (or equivalently,
$\Lambda^\prime(0) \neq 0$).

The regular solution, as well as its radial derivative, vanishes at the singularity.
From this condition and from a slight rearrangement of equation~\eqnref{eqn:u_ODE},

\begin{equation}
	\bvec{u} = \frac{\va^2}{\omega^2 - \omega_A^2} \grad_\perp \Phi \; ,
\end{equation}

\noindent we can deduce that all physical variables (i.e., the magnetic pressure
and both velocity components) vanish at the singularity. On the other hand the
irregular solution possesses a ``buried" logarithmic singularity. By buried we mean
that the pressure variable and its first derivative remain continuous and finite,
whereas its second derivative is singular (i.e., $\Phi_{\rm irr}$ is a $C^1$ function).
In terms of the physical variables, the magnetic pressure is well-behaved, but
both velocity components are divergent,

\begin{eqnarray}
	\Phi_{\rm irr} &=& 1 + \frac{k^2 r_0^2}{2} \, \varepsilon^2 \, \ln\varepsilon + \cdots \; ,
\\
	u_{{\rm irr},r} &=& \frac{k^2 r_0^3}{\Lambda^\prime(0)} \; \ln\varepsilon + \cdots \; ,
\\
	u_{{\rm irr},y} &=& \frac{ikr_0^2}{\Lambda^\prime(0)} \; \frac{1}{\varepsilon} + \cdots
\end{eqnarray}

The presence of the logarithm in the irregular solution is a clear indication that
there are branch cuts in complex radius space. Often clever placement of the branch
cut allows one to essentially ignore their existence. That isn't the case here.
The location of the branch points is a function of the frequency and since the
frequency can be complex---the resonant absorption naturally leads to a loss of
energy from the fast modes through coupling with the Alfv\'en waves, the branch
points can pass through the real radius axis as the imaginary part of the frequency
changes sign. This leads to a branch cut that lies on the real frequency axis and
spans all frequencies for which a resonant Alfv\'en radius exists---see chapter 7
of \cite{Goedbloed:2004}. For our specific Alfv\'en speed profile~\eqnref{eqn:VaProfile},
this Alfv\'en continuum consists of all frequencies, $\left|\omega\right| > mV_0 r_0^{-1}$.


\section{Continuous and Discrete Spectra}
\label{sec:ContSpectra}

The Greens function in spectral space for our ODE~\eqnref{eqn:PhiODE} can be written
in terms of the inner and outer solutions and the dispersion function,

\begin{equation}
	G\left(r; r^\prime, y^\prime \vsep k, \omega\right) = 
		\frac{\Phi\left(r_< \vsep k, \omega\right) \, \Phi\left(r_> \vsep k, \omega\right)}{D(k,\omega)} \; e^{-iky^\prime} \; .
\end{equation}

\noindent To obtain the solution in physical space, one needs to invert both the
spatial and temporal transforms. Traditionally, one inverts the temporal transform
first. In which case, the eigenvalues are complex eigenfrequencies and the axial
wavenumber is a real continuous parameter. To adumbrate, however, this choice leads
to difficulties. Here we make the opposite choice, inverting the spatial transform
first. Thus, the eigenvalues are complex eigenwavenumbers and frequency is a real
continuous parameter.

By analytic contour deformation and the use of the residue theorem, one can verify
that two types of eigenmode are possible. The poles of the Greens function (arising
from the complex zeroes of the dispersion function) correspond to a discrete set
of fast-wave eigenmodes. Any branch points in spectral space, with their associated
branch cuts, lead to continuous spectra. In MHD wave problems with transverse variation
in the Alfv\'en frequency, it is well-known that the continuum of possible Alfv\'en
wave solutions are represented by such a branch cut \citep[i.e.,][]{Uberoi:1972,
Lee:1986, Goedbloed:2004}.

\begin{figure*}
	\epsscale{1.0}
	\plotone{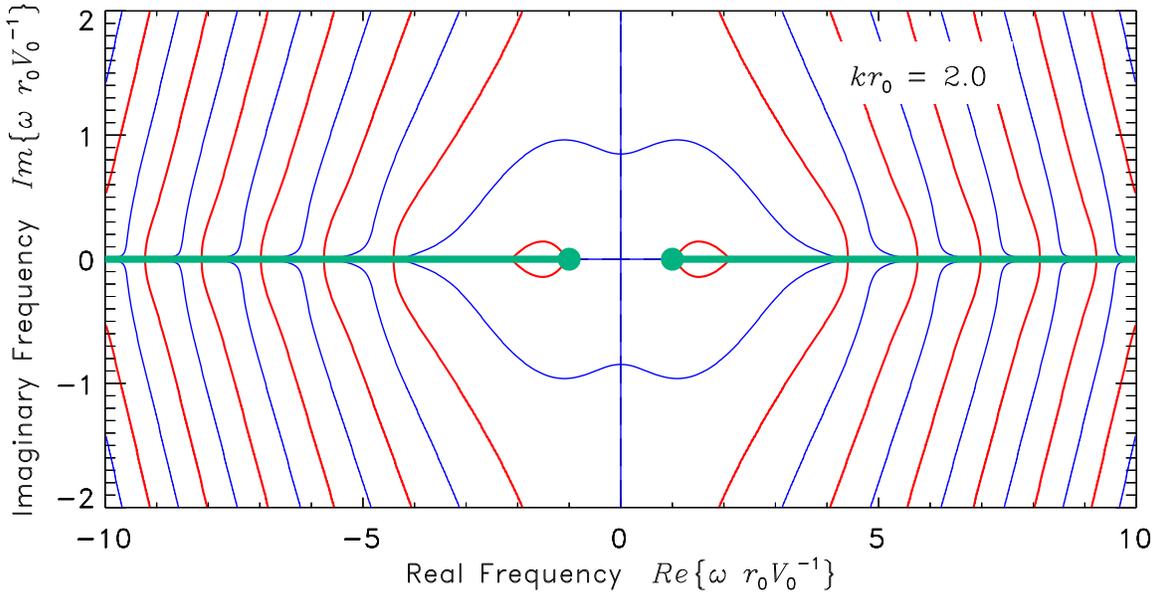}
	\caption{\small The contours in the complex frequency plane where the real
part of the dispersion function $D(k,\omega)$ is zero (red curves) and the imaginary
part is zero (blue curves). The wavenumber used in the illustration is purely real,
$k r_0 = 2$. The azimuthal order is unity, $m=1$. Wherever a red curve and blue
curve cross indicates a complex zero of the dispersion function, and hence the
location of an eigenmode. Notice, that no such crossings occur. The teal lines
mark branch cuts, locations where the dispersion function is discontinuous. The
teal dots indicate the branch points at $\omega =\pm  m V_0 r_0^{-1}$. These branch
cuts are associated with the continuous spectrum of allowed Alfv\'en waves. The
picture illustrates the dispersion function on only the principle Riemann sheet.
The expected fast mode resonances (i.e., zeros of the dispersion function) lie
on nearby Riemann sheets that are reached by analytic continuation of the dispersion
function through the branch cuts.
	\label{fig:AlfvenContinuum}}
\end{figure*}

In our specific problem, the Alfv\'en continuum appears as two branch cuts in
complex frequency space that lie along the real frequency axis for
$\left|\omega\right| > m r_0 V_0^{-1}$. These lines of discontinuity appear because
the irregular solution has a logarithmic singularity and hence there are are a
countable infinity of potential branches or Riemann sheets. For real frequencies
the logarithmic branch point lies on the real radius axis ($\ra$ is real). But,
if the frequency is complex, the radius of the singularity and branch point is
also complex. Further, as the frequency passes across the real axis in the complex
frequency plane, the radial branch point also moves across the real radius axis.
The inner and outer solutions can be viewed as contour integrals along the real
radius axis originating either at the origin or at infinity. Thus, as the radial
branch point moves across the integration contour, there is a discontinuity in
either the inner or outer solution. A careful consideration of the response of
the dispersion function to these discontinuities reveals that there are two branch
points on the real frequency axis at $\omega = \pm m V_0 /r_0$ and branch cuts that
connect these points to $\pm \infty$ along the real axis.  Figure~\ref{fig:AlfvenContinuum}
shows contours of the dispersion function $D(k,\omega)$ as a function of complex
frequency for a real wavenumber $kr_0 = 2$. The red curves indicate where the
real part of the dispersion function is zero and the blue curves are the isocontours
where the imaginary part vanishes. The teal dots indicate the two branch points
and the teal lines display the branch cuts. Note, nowhere do the red and blue
curves cross. Thus, there are no discrete fast eigenmodes that lie on the priniciple
Riemann sheet. Thus, the expected set of fast modes are in this case absorbed
into the continuous spectrum.

The poles associated with the discrete modes do exist, but they appear on nearby
Riemann sheets connected through the branch cuts. It is possible to analytically
continue the solutions through the branch cuts to find these poles. Doing so requires
that one calculate the inner and outer solutions by integrating off the real radius
axis in such a way that the logarithmic singularity never crosses the integration
contour. Unfortunately, this requires that the branch cut in radius space crosses
the real radius axis for frequencies with a negative imaginary part. Thus, one
gains continuity in frequency space at the expense of continuity in radius. Of course
for most applications, this is not a useful trade-off.

\begin{figure*}
	\epsscale{1.0}
	\plotone{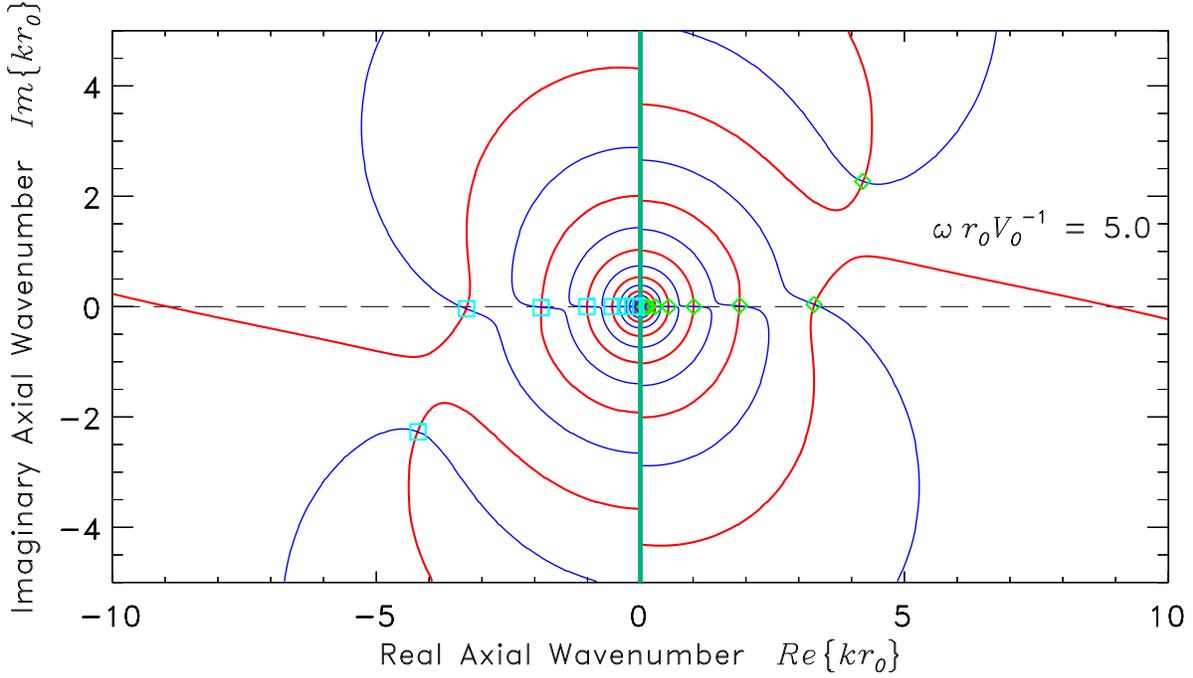}
	\caption{\small The zero contours of the real part (red) and imaginary part
(blue) of the dispersion function in complex wavenumber space. The dispersion function
is illustrated for a real dimensionless frequency, $\omega r_0 V_0^{-1} = 5.0$ and
for a unit azimuthal order, $m=1$. There are many locations where a red curve crosses
a blue curve, indicating a fast wave eigenmode of the arcade. There are two families
of eigenmodes, those indicated with green diamonds are forward-propagating waves
(phase propagates in the positive axial direction), while the aqua squares indicate
the symmetric set of counter-propagating waves. There are a countable infinity of
each type of mode with an accumulation point at the origin. Only a finite number
of modes from each family are indicated in the figure. The imaginary wavenumber
axis is a branch cut associated with the behavior of the eigenfunctions at infinite
radius. This Riemann sheet, the principle sheet, represents only those solutions
which vanish at large radius.
	\label{fig:k-space}}
\end{figure*}

A significantly simpler solution is to invert the spatial transform first and hence
consider complex eigenwavenumbers and real frequencies. Since the location of the
Alfv\'en singularity does not depend on the axial wavenumber, the Alfv\'en continuum
does not insert branch points or other singularities in the complex wavenumber plane.
Figure~\ref{fig:k-space} illustrates the zero contours of the real and imaginary
parts of the dispersion function for a real frequency of $\omega r_0 V_0^{-1} = 5$
and for complex wavenumber. There is a branch cut along the imaginary wavenumber
axis, but it does not arise from the Alfv\'en continuum. Instead it corresponds to
a continuous spectrum of axially evanescent and radially propagating waves akin
to the acoustic jacket modes of helioseismology \citep[see][]{Bogdan:1995}.
Mathematically, this branch cut arises from the behavior of the solution for large
radius. In this limit, the solution behaves like a $K$-modified Bessel function,
$\Phi \propto K_\nu(\sqrt{k^2} \, r)$---see Appendix~\ref{subsec:infinity}. In
order to remain finite at infinite radius, the sign of the square root (i.e., the
branch) must change across the imaginary axis, hence, introducing a branch cut.
This type of continuous spectrum is well-known and appears in a host of simple
problems including the solution in cylindrical coordinates to Poisson's equation
for the electric field generated by a point charge. Note, the red and blue curves
cross each other in this case, indicating discrete modes. There are two families
of modes, those that propagate in the positive $y$-direction (green diamonds) and
those that propagate in the opposite direction (aqua squares). Both families have
an accumulation point of modes at the origin (only a finite number of modes have
been indicated in the figure). The wavenumbers of the two families are anti-symmetric,
with the counter-propagating modes having a wavenumber that is just the negative
of the forward-propagating modes. The weakly damped radial overtones ($n>0$) correspond
to the countable infinity of poles that lie very near the real wavenumber axis.
The strongly damped fundamental mode ($n=0$) appears as two poles (one forward-
and one backward-propagating) in a region of complex wavenumber space well away
from the real wavenumber axis (near $Im\left\{k\right\} = 2.3$).
near


\section{Energy Flux}
\label{sec:EnergyFlux}

The rate at which any given mode loses energy through resonant absorption can be
obtained by considering the flux of wave energy. For a magnetically dominated fluid,
the energy flux $\bvec{F}$ has two components \citep{Bray:1974},

\begin{equation} \label{eqn:EnergyFlux}
	\bvec{F} = \left(\frac{\bvec{B} \cdot \bvec{b}}{4\pi}\right) \bvec{u} - \frac{\bvec{b} \cdot \bvec{u}}{4\pi} \bvec{B} \; .
\end{equation}

\noindent The component of the flux aligned with the background magnetic field
arises from the magnetic tension. While this component is in general a nonzero
function of azimuth, when integrated over azimuth the term vanishes. This occurs
because the transverse components of the perturbed magnetic field $\bvec{b}$
are proportional to $\cos(m\theta)$ whereas the velocity components are proportional
to $\sin(m\theta)$---see equations~\eqnref{eqn:induction} and \eqnref{eqn:mode_form}.
Thus, no net energy is lost or gained by fast waves through the action of this
flux component. The components of the energy-flux that are transverse to the
background field (and result from the magnetic pressure) lack this property.

To ascertain how the energy flux manifests for a single eigenmode, in equation~\eqnref{eqn:EnergyFlux}
we must consider only the real parts of the solution and average in time over
a wave period. The magnetic pressure fluctuation is directly proportional to
$\Phi$---see equation~\eqnref{eqn:pressure_fluctuation},

\begin{equation}
	\Pi = \frac{\bvec{B} \cdot \bvec{b}}{4\pi} = \frac{i}{\omega} \, \frac{B^2}{4\pi} \Phi \; .
\end{equation}

\noindent Therefore denoting the time average by angular brackets, we obtain

\begin{equation}
	\left<\bvec{F}\right> = \frac{B^2}{8\pi\omega} \, Im\left\{\Phi^*(r) \bvec{u}(r)\right\} \, e^{-2\, Im\left\{k\right\}\, y} \, \sin^2\left(m\theta\right) \; ,
\end{equation}

\noindent where we have explicitly included the azimuthal and axial spatial
dependences. Note, because of the time averaging (and the real frequencies),
the energy flux is independent of time. Further, since the eigenwavenumber is
complex, the amplitude of the flux decays in the direction of wave propagation.

The time rate of change of the Alfv\'en wave energy density is given by the
divergence of this energy flux,

\begin{equation}
	\dpar[E]{t} = - \bvec{\nabla} \cdot \left<\bvec{F}\right>
\end{equation}

\noindent It can be demonstrated that the divergence of the energy flux vanishes
everywhere except at the Alfv\'en singularity. To do so, apply the chain rule
to the product $\Phi^*\bvec{u}$ and use equations~\eqnref{eqn:Def_Phi} and
\eqnref{eqn:u_ODE} to obtain,

\begin{equation}
	\bvec{\nabla} \cdot \left<\bvec{F}\right> = \frac{B^2}{8\pi\omega} Im\left\{K^2 \left|\bvec{u}(r)\right|^2
				- \left|\Phi(r)\right|^2 \right\}
				\, e^{-2\, Im\left\{k\right\}\, y} \, \sin^2\left(m\theta\right) = 0 \; .
\end{equation}

\noindent The expression inside the braces is purely real and thus the energy flux is
divergenceless.

This is not a surprising result; the waves are ideal MHD waves. The damping
of the fast modes occurs purely at the singularity due to a transferral of
energy to Alfv\'en waves. Alfv\'en waves are not directly described by our
ODE~\eqnref{eqn:PhiODE} for a variety of reasons: (1) the Alfv\'en waves are
pure tension waves and thus produce no magnetic-pressure fluctuation,
i.e., $\Phi = 0$, and (2) the coupled Alfv\'en waves experience secular
growth and therefore do not have exponential time dependence. This latter
point can be deduced by carefully examining the rate at which the fast mode
pumps energy into the singularity.

As previously stated, the energy flux is divergenceless except at the
singularity itself. However, the energy flux can be shown to have a jump
discontinuity across the singularity. Hence, the divergence of the flux
has a delta function at the singularity. The jump discontinuity arises
from the logarithmic singularity in the irregular solution. As shown in
Appendix~\ref{subsec:interior} the irregular solution has the following
leading order behavior at the singularity,

\begin{eqnarray}
	\Phi_{\rm irr} &=& 1 + \frac{k^2}{2} \, (r-\ra)^2 \, \ln\left(\frac{r-\ra}{r_0}\right) + \cdots \; ,
\\  \label{eqn:ur_sing}
	u_{{\rm irr},r} &=& \frac{k^2 r_0^3}{\Lambda^\prime(0)} \; \ln\left(\frac{r-\ra}{r_0}\right) + \cdots \; ,
\\
	\Lambda^\prime(0) &\equiv& r_0^3 \der[]{r} \left[ \frac{\omega^2 - \oma^2(r)}{\va^2(r)} \right]_{r=\ra} =
		- \frac{r_0^3}{\va^2} \left.\der[\oma^2]{r}\right|_{r=\ra} \; .
\end{eqnarray}

Remember that the analytic continuation of the logarithm to negative arguments
involves an imaginary component with a Heaviside step function, $H$,

\begin{equation}
	\log\left(r-\ra\right) = \log\left|r-\ra\right| + i \pi H\left(\ra - r\right) \; .
\end{equation}

\noindent This step function leads to a discontinuous jump in the energy
flux across the singularity and energy is continually injected into the
singularity. The divergence of the flux has only one term that does not
cancel,

\begin{equation}
	\bvec{\nabla} \cdot \left<\bvec{F}\right> = \frac{I^2 \left|A_{\rm irr}\right|^2}{2\pi\omega} \,
		\frac{Re\left\{k^2\right\} r_0^3}{\Lambda^\prime(0)} \; Im\left\{ \frac{1}{r^2} \der[]{r} \log\left(\frac{r-\ra}{r_0}\right) \right\} \, 
			e^{-2\, Im\left\{k\right\}\, y} \, \sin^2\left(m\theta\right) \; ,
\end{equation}

\noindent The radial dependence has been made obvious by using 
equation~\eqnref{eqn:potential_field} to replace the magnetic field strength
with the constant line-current strength, $B = 2I/r$. The amplitude of the
irregular solution, $A_{\rm irr}$, has units of a frequency and and its
amplitude relative to the regular solution, $A_{\rm irr}/A_{\rm reg}$, is
fixed by the radial boundary conditions. The energy transfer rate has a
delta function for its radial dependence (i.e., arising from the radial
derivative of the step function),

\begin{equation} \label{eqn:AlfvenEnergy}
	\dpar[E]{t} = \frac{I^2 \left|A_{\rm irr}\right|^2}{2\omega} \,
		\frac{Re\left\{k^2\right\} r_0^3}{\Lambda^\prime(0)} \; \frac{\delta(r-\ra)}{\ra^2} \, 
			e^{-2\, Im\left\{k\right\}\, y} \, \sin^2\left(m\theta\right) \; ,
\end{equation}

\begin{figure*}
	\epsscale{1.0}
	\plotone{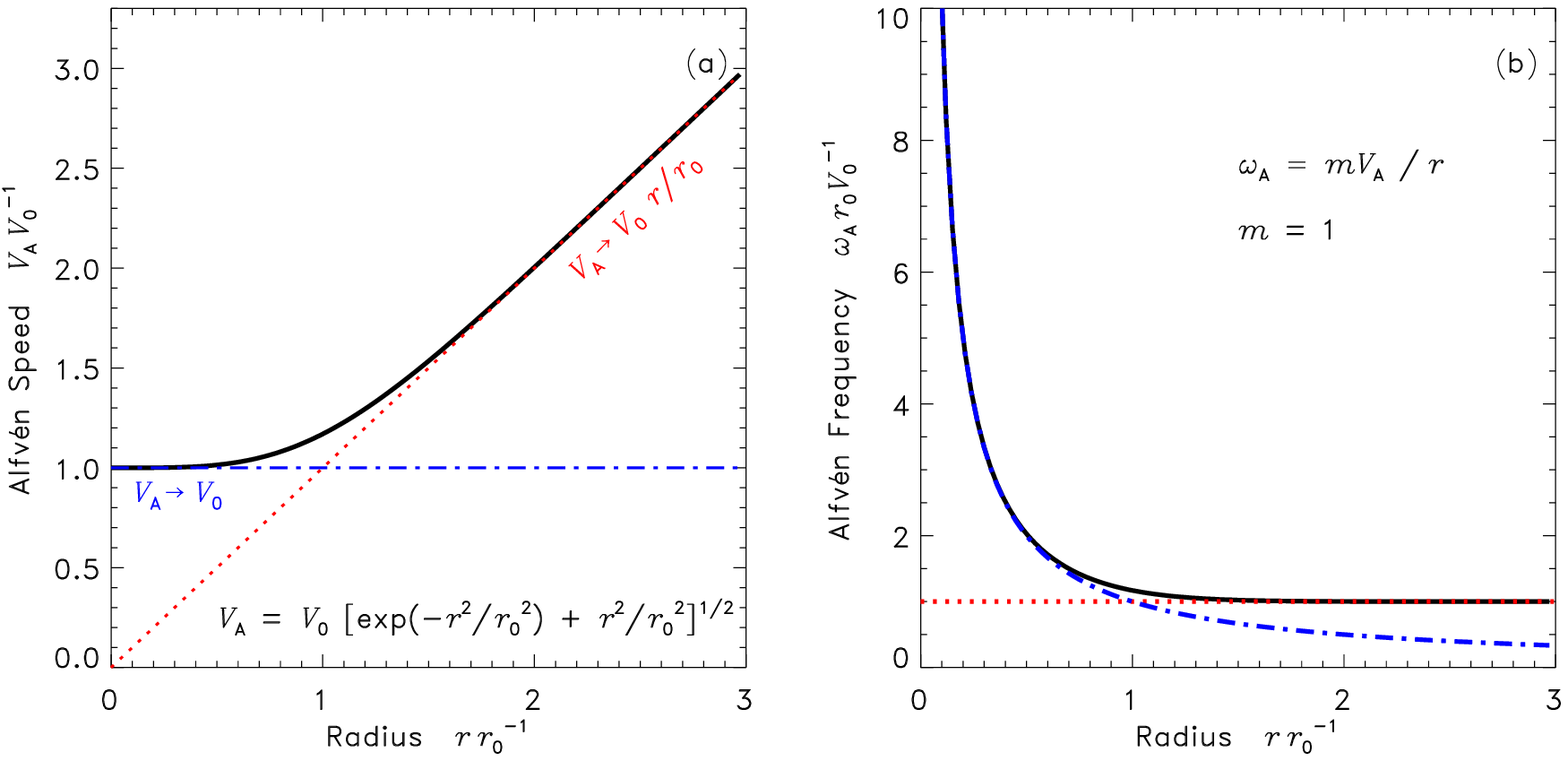}
	\caption{\small (a) The Alfv\'en speed, $\va$, given by equation~\eqnref{eqn:VaProfile}
shown as a function of cylindrical radius. Near the axis, $r<<r_0$, the profile
approaches a constant value, $V_0$ (shown as the blue dot-dashed line). At large
radii, $r>>r_0$, the Alfv\'en speed becomes a linear function of radius (shown as
the red dotted line). (b) The Alfv\'en frequency, $\oma\equiv m\va/r$, displayed
as a function of cylindrical radius for unit azimuthal order, $m=1$. As the radius
becomes large the dimensionless Alfv\'en frequency approaches $m$.
	\label{fig:AlfvenProfile}}
\end{figure*}

For an Alfv\'en frequency that is monotonically decreasing with radius (as we have
here) the derivative of $\Lambda$ is positive, $\Lambda^\prime(0) > 0$ (see
Figure~\ref{fig:AlfvenProfile}). Thus, the eigensolution injects wave energy into
the singularity at a rate that is independent of time.The azimuthally and radially
integrated energy in the Alfv\'en waves grows at a linear rate, $E = E_0\left(1+t/\tau\right)$,
with $E_0$ being the energy at time $t=0$ and $\tau^{-1}$ given by integrating
the right-hand side of equation~\eqnref{eqn:AlfvenEnergy} and divided by $E_0$,

\begin{equation}
	\tau^{-1} = \frac{\pi I^2 \left|A_{\rm irr}\right|^2}{4\omega E_0} \,
		\frac{Re\left\{k^2\right\} r_0^3}{\Lambda^\prime(0) \ra} \, 
			e^{-2\, Im\left\{k\right\}\, y} \; .
\end{equation}

This linear growth of the Alfv\'en wave energy density can be viewed as the root
necessity for the Alfv\'en continuum. The temporal behavior of the Alfv\'en waves cannot
be expressed with a discrete set of exponentials, hence they must be represented as
an integral that adds up the contribution of a continuum of exponential forms. This
linear combination has the relative phase relations required to build a polynomial
in time. This procedure is exactly what continuous spectra from a branch cut often
represent.




\end{document}